\documentclass[prb,superscriptaddress,showpacs,longbibliography,reprint]{revtex4-2}

\usepackage[colorlinks=true,linkcolor=blue,urlcolor=blue,citecolor=blue,pdfusetitle]{hyperref}
\usepackage{color}
\usepackage[utf8]{inputenc}
\usepackage[usenames,dvipsnames]{xcolor}
\usepackage{amsmath}
\usepackage{amssymb}
\usepackage{graphicx}
\graphicspath{{graphics/}}
\usepackage{epsfig}
\usepackage{dcolumn}% Align table columns on decimal point
\usepackage{bm}% bold math
\usepackage{mathrsfs}
\usepackage{multirow}
\usepackage[all]{xy}
\usepackage{pbox}
\usepackage{lipsum}
\usepackage{verbatim}
\usepackage{braket}
\usepackage{dsfont}
\usepackage{array}
\usepackage{makecell}
\usepackage{tabularx}
\usepackage{isotope}
\usepackage{xr}
\usepackage[normalem]{ulem}

\usepackage[dvipsnames]{xcolor}

\begin{document}

\title{Nonequilibrium transition between dissipative time crystals}

\author{Albert Cabot}
\affiliation{Institut für Theoretische Physik, Eberhard Karls Universität Tübingen, Auf der Morgenstelle 14, 72076 Tübingen, Germany}
\author{Gian Luca Giorgi}
\affiliation{IFISC (UIB-CSIC), Instituto de F\'isica Interdisciplinar y Sistemas Complejos, Palma de Mallorca, Spain}
\author{Roberta Zambrini}
\affiliation{IFISC (UIB-CSIC), Instituto de F\'isica Interdisciplinar y Sistemas Complejos, Palma de Mallorca, Spain}

\begin{abstract}
We show a dissipative phase transition in a driven nonlinear quantum oscillator in which a discrete time-translation symmetry is spontaneously broken in two different ways. The corresponding regimes display either discrete or incommensurate time-crystal order, which we analyze numerically and analytically beyond the classical limit addressing observable dynamics, phenomenology in different (laboratory and rotating) frames, Liouvillian spectral features, and quantum fluctuations.
Via an effective semiclassical description, we show that phase diffusion dominates in the incommensurate time crystal (or continuous time crystal in the rotating frame), which manifests as a band of eigenmodes with a lifetime growing linearly with the mean-field excitation number. Instead, in the discrete time crystal phase, the leading fluctuation process corresponds to quantum activation with a single mode that has an exponentially growing lifetime. Interestingly, the transition between these two regimes manifests itself already in the quantum regime as a spectral singularity, namely as an exceptional point mediating between phase diffusion and quantum activation. Finally, we discuss this transition between different time-crystal orders in the context of synchronization phenomena.
\end{abstract}

\maketitle

\section{introduction}

Driven-dissipative quantum systems display genuine non-equilibrium phases and dissipative phase transitions emerging from the interplay of driving, dissipation, and interactions. Their characterization, critical dynamics, and universal features is the focus of a large body of theoretical \cite{Haken2006,Lugiato2015,Diehl2008,Diehl2010,Lesanovsky2013,Lee2013b,Jin2013,Marcuzzi2014,Carmichael2015,Marcuzzi2016,Casteels2016,Jin2016,Bartolo2016,Casteels2017,Hwang2018,Rota2017,Foss-Feig2017,Rota2019,Jager2019} and experimental \cite{Carr2013,Fitzpatrick2017,Fink2018,Li2022,Benary2022} work. In recent years, intense activity has focused on non-equilibrium phases of matter in which time-translation symmetry is spontaneously broken, the so-called time crystals. 

The existence of time crystals was initially discussed for closed Hamiltonian systems \cite{Wilczek2012}, however, it was later shown that they cannot emerge in equilibrium systems with short-range interactions \cite{Bruno2013,Bruno2013b,Nozieres2013,Watanabe2015}.  Attempts have been made to circumvent the possibility of observing time crystals in Hamiltonian systems which are based either on long-range multispin interaction \cite{Kozin_2019} or
interacting gauge fields \cite{PhysRevLett.123.250402}, although these approaches have also been criticized \cite{khemani2020comment,PhysRevLett.124.178901}.  Far from equilibrium, time crystals have been reported both in driven-Hamiltonian \cite{Khemani2016,Else2016,Yao2017,Sacha2018,Else2020,Zalatel2023} and driven-dissipative systems \cite{Iemini2018,Sacha2018,Else2020}. In driven-dissipative scenarios,  time-translation symmetry  can appear due to different mechanisms, which provoke the emergence of nonequivalent forms of time crystals. In time-independent systems (or in an appropriate rotating frame) this symmetry can be broken continuously, leading to continuous time crystals \cite{Iemini2018,Tucker2018,Lledo2019,Lledo2020,Buonaiuto2021,Carollo2022,Seibold2020,Buca2019,Buca2019b,Booker2020,Hajdusek2022,Prazeres2021,Piccitto2021,Hajdusek2022,Buca2022,Colella2022,Krishna2023,Nie2023,Cabot2023,Nakanishi2023,Gao2023,Kosior2023,Cabot2024}. In time-dependent periodic systems this symmetry can be broken discretely, e.g. as a rigid subharmonic response to an external periodic forcing, which are known as discrete (dissipative) time crystals \cite{Gong2018,Wang2018,Dykman2018,Gambetta2019,Lazarides2020,Riera2020,Zhu2019,Chinzei2020,Tuquero2022,Chitra2015,Cabot2022b,Nie2023}. Time-dependent scenarios can also lead to the emergence of stable incommensurate responses. In the context of driven Hamiltonian systems, these are generally referred to as time quasicrystals \cite{Giergiel2019,Pizzi2019,Zhao2019,Yang2021}, while in driven-dissipative scenarios the name { incommensurate time crystal} is also used \cite{Cosme2019,Chinzei2020,Homann2020,Skulte2021,Nie2023,Cosme2023}. Signatures of these different time-crystalline orders have been recently reported in various experimental platforms, studying both driven-Hamiltonian systems \cite{Zhang2017,Choi2017,Autti2018,Randall2021,Frey2022}, driven-dissipative atomic systems \cite{Kessler2021,Kongkhambut2021,Kongkhambut2022} and nonlinear optical cavities \cite{Taheri2022}.

The eigenspectrum of the dynamical map describing an open quantum system provides a powerful tool to analyze the emergence of symmetry broken phases and dissipative phase transitions (DPT) \cite{Kessler2012,Albert2014,Minganti2018}, as well as dynamical phenomena like metastability \cite{Macieszczak2016,Cabot2022b} or quantum synchronization \cite{Galve2017,Giorgi2019}. DPTs can be signaled by a Liouvillian gap closure in the thermodynamic \cite{Kessler2012} and infinite-excitation limits \cite{Minganti2018,Carmichael2015}. In continuous time crystals a set of Liouvillian eigenvalues becomes purely imaginary in these limits \cite{Iemini2018}. While in discrete (dissipative) time crystals, one can find the signatures in the Floquet map spectrum, in which a set of eigenvalues acquire a unit absolute value \cite{Gong2018,Riera2020,Cabot2022b}. The properties of the dominant eigenvalues explain the distinctive emergent behavior characterized by non-decaying oscillatory responses either with a continuous period \cite{Iemini2018,Lledo2019,Buca2019} or a rigid subharmonic response \cite{Gong2018,Riera2020,Cabot2022b}. Therefore,  for a deeper understanding of these emergent phenomena, it is crucial to understand the behavior of the leading eigenvalues of the dynamical generator. This analysis can shed light on the dominant fluctuation processes that affect such phenomena and facilitates the characterization and classification of nonequilibrium phases and phase transitions.

In this work, we consider the quantum van der Pol oscillator (QvdP) with squeezed drive \cite{Sonar2018,Kato2019,Mok2020,Kato2021,Cabot2021}. This is a paradigmatic system in the study of quantum synchronization \cite{Galve2017}. It allows for the study of synchronization phenomena from the classical to the quantum regime \cite{Lee2013,Walter2014,Lee2014,Walter2015}. In the presence of a squeezed forcing, this system exhibits a nonequilibrium transition between a phase characterized by a stable incommensurate response (incommensurate time crystal) and a phase characterized by a stable subharmonic response (discrete time crystal). Interestingly, in this setup, the explicit time dependence of the model can be eliminated by moving to the driving co-rotating frame. In such a frame,  we demonstrate that the incommensurate time crystal manifests as a continuous time crystal 
and the discrete time crystal manifests as a parity-broken stationary phase. In Table  \ref{tab:table1}, we resume the different kinds of emerging mechanisms depending on the frame and on the system parameters.  This provides an intriguing scenario in which to address the interrelation between different symmetry-broken phases and fundamental questions such as: (i) how different time-crystal orders shape the spectral properties and fluctuation processes occurring in the same degrees of freedom; and (ii) how these properties and processes reorganize themselves through a nonequilibrium transition or DPT. In Sections \ref{Sec_continuous} and \ref{Sec_discrete}  we address the question (i), analyzing both the Liouvillian spectrum and system observables. We show that a semiclassical effective phase model allows us to identify the dominant fluctuation processes of each symmetry-broken regime and to explain the asymptotic behavior of the Liouvillian spectrum. In Sec. \ref{Sec_ep} we address the question (ii), showing that a spectral singularity, namely an exceptional point (EP) \cite{Heiss2012,Minganti2019}, mediates in between these two regimes. This provides a signature of the nonequilibrium transition far from the infinite-excitation limit and marks the point at which the dominant eigenvalues change qualitatively their behavior. We present our conclusions in Sec. \ref{sec_conclusions}, where we discuss the prospect of observing similar phenomena in general synchronization scenarios.

\section{Model and symmetries}

The squeezed QvdP oscillator consists of a bosonic mode coherently driven by a two-boson term (the squeezing) and subject to linear amplification and two-boson dissipation. The model is described by the following master equation in the laboratory frame ($\hbar=1$) \cite{Sonar2018,Cabot2021}:
\begin{equation}
\begin{split}
\partial_t \hat{\rho}_\mathrm{L}&=-i[\Hat{H}_\mathrm{L}(t),\hat{\rho}_\mathrm{L}]+\frac{\gamma_1}{2}\mathcal{D}[\hat{a}^\dagger]\hat{\rho}_\mathrm{L}+\frac{\gamma_2}{2}\mathcal{D}[\hat{a}^2]\hat{\rho}_\mathrm{L}\\
&=\mathcal{L}_\mathrm{L}(t)\hat{\rho}_\mathrm{L},
\end{split}
\end{equation}
with
\begin{equation}\label{eq:HL}
\hat{H}_\mathrm{L}(t)=\omega_0\hat{a}^\dagger\hat{a}+i\eta(\hat{a}^2 e^{i2\omega_\mathrm{s} t}-\hat{a}^{\dagger 2} e^{-i2\omega_\mathrm{s} t}),    
\end{equation}
where $\omega_0$ is the frequency of the mode, $\eta$ and $2\omega_\mathrm{s}$ the squeezing strength and its frequency, $\gamma_1$ the amplification rate and $\gamma_2$ the two-boson dissipation rate. The Lindblad dissipator is here defined as $\mathcal{D}[\hat{L}]\hat{\rho}=2\hat{L}\hat{\rho}\hat{L}^\dagger-\hat{L}^\dagger \hat{L}\hat{\rho}-\hat{\rho}\hat{L}^\dagger \hat{L}$. The master equation displays a discrete time-translation symmetry with period $T=\pi/\omega_s$, i.e. $\mathcal{L}_L(t+T)=\mathcal{L}_L(t)$. This model also displays a parity symmetry, as revealed by its invariance under the transformation $\hat{a}(\hat{a}^\dagger)\to -\hat{a}(-\hat{a}^\dagger)$. This can be formally expressed as $[\mathcal{Z}_2,\mathcal{L}_\mathrm{L}(t)]=0$, where  $\mathcal{Z}_2(\cdot)=e^{-i\pi\hat{a}^\dagger\hat{a}}\cdot e^{i\pi\hat{a}^\dagger\hat{a}}$ is the parity superoperator \cite{Minganti2018,Albert2014}.

{\it Rotating frame. --} The explicit time-dependence of this Liouvillian can be conveniently eliminated in a rotating frame, as defined by a time-dependent unitary transformation: $\hat{U}_t=\exp(-i\omega_\mathrm{s} \hat{a}^\dagger \hat{a} t)$. In this frame, the Hamiltonian reads
\begin{equation}
\hat{H}=\Delta\hat{a}^\dagger\hat{a}+i\eta(\hat{a}^2 -\hat{a}^{\dagger 2}),
\end{equation}
with the detuning defined as $\Delta=\omega_0-\omega_\mathrm{s}$. This transformation does not affect the other terms in the master equation, which now reads:
\begin{equation}
\begin{split}
\partial_t \hat{\rho}&=-i[\Hat{H},\hat{\rho}]+\frac{\gamma_1}{2}\mathcal{D}[\hat{a}^\dagger]\hat{\rho}+\frac{\gamma_2}{2}\mathcal{D}[\hat{a}^2]\hat{\rho}
=\mathcal{L}\hat{\rho},
\end{split}
\end{equation}
and which describes the dynamics of the rotating state:
\begin{equation}\label{lab_to_rot}
\hat{\rho}(t)=\hat{U}_t^\dagger \hat{\rho}_\mathrm{L}(t)\hat{U}_t.
\end{equation}

In the following,  we first analyze the dynamics in the rotating frame, which is easier to compute. Then, we examine its counterpart in the laboratory frame through the unitary (\ref{lab_to_rot}).  As we show below, the spontaneous breaking of parity symmetry is equivalent (in this system) to the emergence of a stable subharmonic response in the laboratory frame. On the other hand, the continuous breaking of time-translation symmetry in the rotating frame is identified as the emergence of non-decaying oscillations \cite{Iemini2018} whose fundamental frequency varies continuously with the system parameters and which manifest as an incommensurate stable response when we move to the laboratory frame (see Table \ref{tab:table1}).

\begin{table}
\caption{\label{tab:table1} Reference frame and emerging phases. The parameter $\eta$ measures the driving strength and is defined in Eq. (\ref{eq:HL}), while the critical value $\eta_c$ is defined in Eq. (\ref{eq:etac}). }
\begin{ruledtabular}
\begin{tabular}{lcc}
& Laboratory frame & Rotating frame\\
\hline
$\eta<\eta_\mathrm{c}$ & Incommensurate TC\footnote{See also Refs.  \cite{Cosme2019,Chinzei2020,Homann2020,Skulte2021,Nie2023,Cosme2023}.}& Continuous TC\footnote{See also Refs.  \cite{Iemini2018,Tucker2018,Lledo2019,Lledo2020,Buonaiuto2021,Carollo2022,Seibold2020,Buca2019,Buca2019b,Booker2020,Hajdusek2022,Prazeres2021,Piccitto2021,Hajdusek2022,Buca2022,Colella2022,Krishna2023,Nie2023,Cabot2023,Nakanishi2023,Gao2023,Kosior2023,Cabot2024}.}\\
$\eta>\eta_\mathrm{c}$& Discrete TC\footnote{See also Refs. \cite{Gong2018,Wang2018,Dykman2018,Gambetta2019,Lazarides2020,Riera2020,Zhu2019,Chinzei2020,Tuquero2022,Chitra2015,Cabot2022b,Nie2023}.} & Parity ($\mathcal{Z}_2$) symmetry breaking\\
\end{tabular}
\end{ruledtabular}
\end{table}
\section{Classical dynamics}\label{Sec_MF}

The first step in the analysis is to consider the mean-field description, where quantum correlations and fluctuations are completely neglected. This provides us with a basic description of the system dynamics, from which the bifurcation diagram can be retrieved. In QvdP systems this description becomes accurate for large bosonic occupation numbers, which can be achieved when the amplification rate is much larger than the non-linear dissipation: $\gamma_1/\gamma_2\gg1$ \cite{Lee2013}.

\begin{figure}[t!]
 \centering
 \includegraphics[width=1\columnwidth]{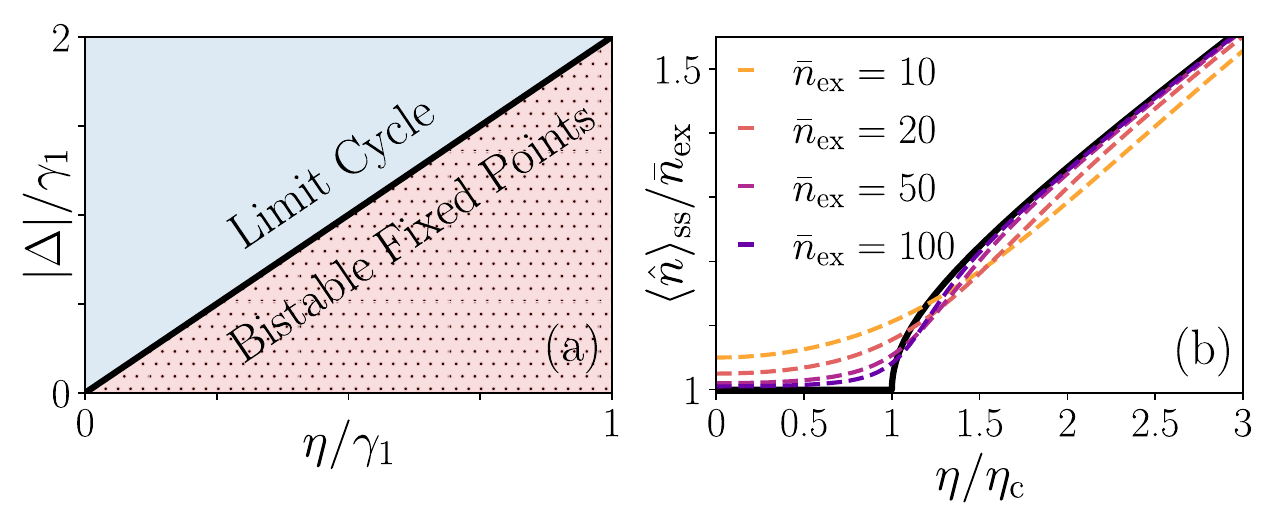}
 \caption{{\bf Bifurcation diagram.} (a) In the rotating frame the mean-field dynamical system displays a stable limit cycle for $\eta<\eta_\mathrm{c}=|\Delta|/2$ or two stable fixed points for $\eta>\eta_\mathrm{c}$. (b) In color dashed lines: stationary occupation number for the full quantum model, $\Delta/\gamma_1=0.1$ and different $\bar{n}_\mathrm{ex}$. Black-solid line: mean-field results (averaged over a period). }
 \label{fig_diagram}
\end{figure}

{\it Mean-field bifurcation diagram in the rotating frame. --}  The mean-field equations of motion are obtained from  those of $\langle \hat{a}\rangle$ factorizing higher order moments $\langle \hat{a}^\dagger \hat{a} \hat{a}\rangle\to\langle\hat{a}^\dagger\rangle\langle\hat{a}\rangle\langle\hat{a}\rangle$ and identifying the complex amplitude as $\alpha=\langle\hat{a}\rangle$. This yields the following non-linear equation for the complex amplitude:
\begin{equation}\label{MF_eq}
\dot{\alpha}=-i\Delta \alpha+\frac{\gamma_1}{2}\alpha-\gamma_2|\alpha|^2\alpha-2\eta\alpha^*.     
\end{equation}
The mean-field dynamical regimes are particularly simple because they depend only on the relationship between the detuning, $\Delta$, and the squeezing strength, $\eta$.  This becomes apparent after rewriting the amplitude in terms of the intensity $N$ and the phase $\phi$: $\alpha=\sqrt{N}e^{i\phi}$. The equations of motion for the new variables are:
\begin{equation}\label{MF_eq_polar}
\begin{split}
\dot{N}&=\gamma_1 N-2\gamma_2N^2-4\eta N\cos(2\phi),\\    
\dot{\phi}&=-\Delta+2\eta\sin(2\phi).    
\end{split}
\end{equation}
These equations display a limit cycle for $\eta<\eta_\mathrm{c}$ and two stable fixed points for $\eta\geq\eta_\mathrm{c}$ \cite{Sonar2018,Kato2021}, with
\begin{equation}\label{eq:etac}
\eta_\mathrm{c}=\frac{|\Delta|}{2}.
\end{equation}
The bifurcation diagram is summarized graphically in Fig. \ref{fig_diagram} (a).  The limit-cycle regime is characterized by an average (over a period) intensity
\begin{equation}\label{MF_LC_intensity}
\bar{N}=\frac{\gamma_1}{2\gamma_2}    
\end{equation}
and a fundamental frequency given by
\begin{equation}\label{freq_LC}
\Omega=\sqrt{\Delta^2-4\eta^2}.    
\end{equation}
The bistable regime is made of two fixed points given by 
\begin{equation}
\alpha_\pm=\pm \sqrt{N_\mathrm{ss}} e^{i\phi_\mathrm{ss}},
\end{equation}
with \cite{Sonar2018}
\begin{equation}\label{MF_solution}
\begin{split}
&N_\mathrm{ss}=\frac{\gamma_1}{2\gamma_2}+\frac{1}{\gamma_2}\sqrt{4\eta^2-\Delta^2}, \\
&2\phi_{\mathrm{ss}}=\pi-\sin^{-1}\frac{\Delta}{2\eta},
\end{split}
\end{equation}
and thus related by a parity transformation, i.e. a $\pi$-phase. The transition corresponds to an infinite period bifurcation in which a limit cycle disappears and gives birth to two saddle nodes  \cite{Kato2021}. This contrasts the more familiar (supercritical) Hopf bifurcation in which the period is finite and the amplitude of the cycle gradually builds up after the transition \cite{Strogatz2018}. At the mean-field level, the limit-cycle breaks continuously the time-translation symmetry, while the fixed points constitute parity-broken solutions.

{\it Mean-field dynamics in the laboratory frame. --} In the laboratory frame, the amplitude of the system acquires the time-dependent phase factor $e^{-i\omega_\mathrm{s}t}$. Since $\omega_\mathrm{s}$ and $\Omega$ are generally incommensurate the limit cycle, whose oscillation contains multiple harmonics of $\Omega$,  manifests as a quasiperiodic oscillation in the laboratory frame. On the other hand, the bistable regime for $\eta>\eta_\mathrm{c}$ becomes a phase characterized by a subharmonic response, $\alpha_\pm(t)=\pm \sqrt{N_\mathrm{ss}} e^{i\phi_\mathrm{ss}-i\omega_\mathrm{s}t}$, breaking the discrete time-translation symmetry with an oscillation at half of the driving frequency. In Table \ref{tab:table1} we summarize the correspondence between the different phases in the two frames, indicating other setups in which similar non-equilibrium phases have been observed. Notice that many of the continuous time crystals reported in the literature also manifest as inconmensurate time crystals in the corresponding laboratory frame.

{\it Quantum to classical crossover. --} The bifurcation diagram is independent of the ratio $\gamma_1/\gamma_2$, which only controls the amplitude of the solutions, i.e., how excited the system is. This motivates the definition of the following parameter:
\begin{equation}
\bar{n}_\mathrm{ex}\equiv \bar{N}=\frac{\gamma_1}{2\gamma_2}.
\end{equation}
At the mean-field level, this is just the scaling factor that controls the amplitude of the oscillations; $\bar{n}_\mathrm{ex}$ exactly corresponds to the mean-field intensity of the limit cycle [Eq. (\ref{MF_LC_intensity})], while it also gives the dominant contribution to the intensity of the bistable fixed points near the bifurcation [Eq. (\ref{MF_solution})]. In fact, the order of magnitude of the boson number in the quantum oscillator is well captured by this parameter, so we will refer to it as the {\it mean-field excitation number}. It can also be used to explore the transition between the quantum (small $\bar{n}_\mathrm{ex}$) and classical (large $\bar{n}_\mathrm{ex}$) limits of the system \cite{Lee2013}. This is shown in Fig. \ref{fig_diagram} (b), where we compare the mean-field solution with the stationary boson number of the quantum system, obtained by numerical integration of the master equation, as $\bar{n}_\mathrm{ex}$ varies. We can clearly see that for large $\bar{n}_\mathrm{ex}$ the quantum results tend to collapse on the mean-field results (averaged over a period). 
In order to fully characterize the emergence of these non-equilibrium regimes, one must go beyond the stationary state quantities and study the dynamics of the observables and the distinctive features of the Liouvillian spectrum. Equivalently, it is necessary to identify the leading fluctuation processes accompanying the mean-field dynamics and how they behave with $\bar{n}_\mathrm{ex}$.

\section{Spontaneous symmetry breaking in the rotating frame}\label{Sec_continuous}

In this section, we address the full quantum model in the rotating frame and analyze the emergence of the different symmetry-broken phases as $\bar{n}_\mathrm{ex}\to\infty$. We begin addressing the breaking of continuous time-translation symmetry for $\eta<\eta_\mathrm{c}$, and later on we address the spontaneous parity breaking occurring for $\eta>\eta_\mathrm{c}$. In both cases, we first analyze how they manifest in the dynamics of observables and the Liouvillian spectrum and afterwards we present a semiclassical fluctuation model in which mean-field dynamics is supplemented by classical white noise. This enables us to understand the behavior of the dominant Liouvillian eigenvalues in these symmetry-breaking regimes and unveil the dominant fluctuation processes.

\subsection{Continuous time crystal: full quantum model}

In driven-dissipative quantum systems, continuous time-translation symmetry breaking occurs when there are observables for which there emerge non-decaying persistent oscillations that evolve in time according to some function \cite{Iemini2018,Lledo2020}: 
\begin{equation}\label{timecrystals_Continuous}
f(\tau)=\lim_{t\to\infty}\lim_{\bar{n}_\mathrm{ex}\to\infty}\text{Tr}[\hat{O}\hat{\rho}(t+\tau)],
\end{equation}
Here $f(\tau)$ is a periodic function whose period varies {\it continuously} with the system parameters. 
Interestingly, the two limits in Eq. (\ref{timecrystals_Continuous}) generally do not commute, since for finite sizes (or excitation number, as in our case) the master equation usually has a unique time-independent stationary state that respects the Liouvillan symmetry and is then incompatible with a symmetry-broken regime.

\begin{figure}[t!]
 \centering
 \includegraphics[width=1\columnwidth]{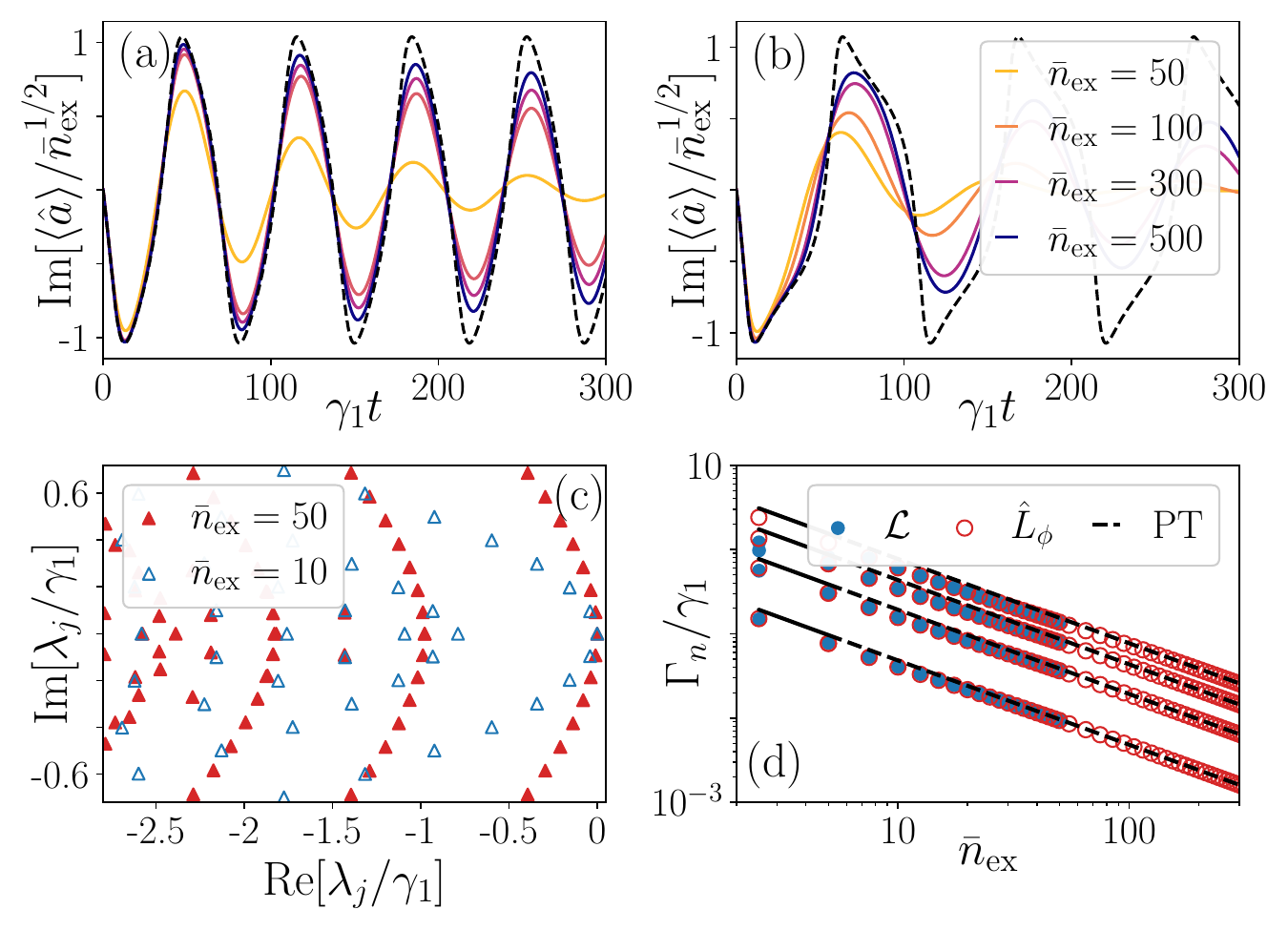}
 \caption{{\bf Signatures of continuous time-translation symmetry breaking.} (a), (b) Color-solid lines: imaginary part of the amplitude dynamics for different $\bar{n}_\mathrm{ex}$, $\Delta/\gamma_1=0.1$ and (a) $\eta/\eta_\mathrm{c}=0.4$ or (b) $\eta/\eta_\mathrm{c}=0.8$. Initial condition: coherent state of amplitude $\sqrt{\bar{n}_\mathrm{ex}}$. In black-dashed lines: mean-field dynamics. (c) Leading eigenvalues of the Liouvillian for $\eta/\eta_\mathrm{c}=0.4$, $\Delta/\gamma_1=0.1$ and $\bar{n}_\mathrm{ex}=10$ (blue-empty triangles) or $\bar{n}_\mathrm{ex}=50$ (red-full triangles). (d) Blue-full circles: first four leading decay rates of the Liouvillian $\mathcal{L}$. Red-empty circles:  first four leading decay rates of the phase Fokker-Planck operator $\hat{L}_\phi$. Black-dashed lines: perturbation theory results for $\hat{L}_\phi$ [Eq. (\ref{phase_eigenvalues})]. Perturbative results predict a scaling $\Gamma_n/\gamma_1=c_n/\bar{n}_\mathrm{ex}$, with $c_1=0.48$, $c_2=1.93$, $c_3=4.34$ and $c_4=7.72$. Parameters for panel (d): $\eta/\eta_\mathrm{c}=0.4$ and $\Delta/\gamma_1=0.1$.}
 \label{fig_continuous}
\end{figure}

{\it Signatures in the dynamics of expectation values. --}  The progressive emergence of the limit-cycle solution can be observed in the dynamics of the mean amplitude, as shown in Fig. \ref{fig_continuous} (a) and (b). Here, we present $\text{Im}[\langle \hat{a}(t)\rangle/\bar{n}_\mathrm{ex}^{1/2}]$ increasing $\bar{n}_\mathrm{ex}$ and taking as initial condition a coherent state of equal amplitude $\bar{n}_\mathrm{ex}^{1/2}$ to the classical limit cycle [cf. Eq. (\ref{MF_LC_intensity})]. In this figure, the mean-field self-sustained oscillation is also shown (black-dashed line) for $\eta/\eta_\mathrm{c}=0.04$ (a), and  $\eta/\eta_\mathrm{c}=0.8$ (b). As the squeezing strength is increased, more harmonics are involved in the oscillation. In both cases, we observe how the lifetime of the oscillations increases with $\bar{n}_\mathrm{ex}$. Moreover, the quantum oscillations increasingly overlap with the mean-field ones for larger $\bar{n}_\mathrm{ex}$, indicating the emergence of non-decaying oscillations \cite{Iemini2018}.

{\it Signatures in the Liouvillian spectrum. --} The Liouvillian spectrum provides a way to systematically analyze the dynamics of a driven-dissipative system. The eigenspectrum is composed of the right and left Liouvillian eigenmatrices together with their eigenvalues:
\begin{equation}
\mathcal{L}\hat{r}_j=\lambda_j\hat{r}_j,\quad  \mathcal{L}^\dagger\hat{l}_j=\lambda^*_j\hat{l}_j,
\end{equation}
where, if diagonalizable, these form a biorthogonal basis Tr$[\hat{l}_j^{\dagger} \hat{r}_k]=\delta_{jk}$. Assuming the presence of a steady state $\hat{\rho}_\mathrm{ss}$ (this is always the case in finite dimensions, while it needs to be checked in the case of infinite-dimensional Hilbert space \cite{evans1977generators,baumgartner2008analysis}), this corresponds to $\lambda_0=0$, $\hat{\rho}_\mathrm{ss}=\hat{r}_0/\text{Tr}[\hat{r}_0]$ and $\hat{l}_0=\mathbb{I}$. The rest of the eigenvalues have a non-positive real part which can be used to order them: $\text{Re}[\lambda_1]\geq \text{Re}[\lambda_2]\geq \text{Re}[\lambda_3]\dots$ The eigenmatrices can be used to decompose the dynamics of the state of the system as
\begin{equation}
\hat{\rho}(t)=\hat{\rho}_\mathrm{ss}+\sum_{j\geq 1}\text{Tr}[\hat{l}^\dagger_j\hat{\rho}(0)]\hat{r}_j e^{\lambda_j t},    
\end{equation}
which shows that the timescales in the dynamics can be generally understood from $\lambda_j$ (this would not be necessarily the case in the presence of collective phenomena such as skin effects \cite{haga2021liouvillian}). Defining their real and imaginary part as their frequency and decay rate, respectively:
\begin{equation}
\varepsilon_j=\text{Im}[\lambda_j],\quad     \Gamma_j=-\text{Re}[\lambda_j],
\end{equation}
we see that the criterium for non-decaying oscillations is that $\Gamma_j=0$ and $\varepsilon_j\neq0$ for some $j$. This can emerge in some scaling limit \cite{Iemini2018}, e.g.:
\begin{equation}
\lim_{\bar{n}_\mathrm{ex}\to\infty} \Gamma_j=0,\quad    \lim_{\bar{n}_\mathrm{ex}\to\infty} \varepsilon_j\neq 0. 
\end{equation}
%. 

{\it Bands of eigenmodes. --} In Fig. \ref{fig_continuous} (c) and (d) we analyze the behavior of the leading Liouvillian eigenmodes with $\bar{n}_\mathrm{ex}$ and $\eta<\eta_\mathrm{c}$. In panel (c) we show the leading eigenvalues for $\eta/\eta_\mathrm{c}=0.4$, $\Delta/\gamma_1=0.1$ and  $\bar{n}_\mathrm{ex}=10$ (blue empty triangles) and $\bar{n}_\mathrm{ex}=50$ (red triangles). In both cases,  there appear many eigenvalues with approximately the same frequency but different decay rates. We refer to the ones with the smallest decay rate for each frequency as {\it the fundamental band} of modes, while those with approximately the same frequency and larger decay rates form the subsequent {\it higher-order bands} of modes. The band structure actually disappears for $\eta>\eta_\mathrm{c}$, in the discrete time-crystal regime. The decay rates do establish the notion of the continuous time crystal: while the ones of the fundamental band diminish with $\bar{n}_\mathrm{ex}$,  the ones of the rest of the bands display a different behavior saturating to $\sim\gamma_1$ (see App. \ref{app_continuous}).

{\it Asymptotic behavior of the fundamental band. --} In  Fig. \ref{fig_continuous} (d) we focus on the behavior of the decay rates of the fundamental with $\bar{n}_\mathrm{ex}$ (blue dots). We focus on the four smallest rates as their asymptotic behavior is more evident for finite $\bar{n}_\mathrm{ex}$. As we can see they diminish with $\bar{n}_\mathrm{ex}$ and asymptotically approach the black-dashed lines, which scale linearly with the inverse of the mean-field excitation number, i.e. $\propto 1/\bar{n}_\mathrm{ex}$.  This scaling behavior is the signature of continuous time-translation symmetry breaking in the infinite-excitation limit \cite{Iemini2018}, while the inverse proportionality with system size has also been reported for spin systems \cite{Buonaiuto2021,Nakanishi2023}. Moreover, notice that their imaginary parts show a band structure, as they tend to the mean-field frequency or multiples of it (see App. \ref{app_continuous}). Thus, the fundamental band is behind the emergence of persistent oscillations observed in Fig. \ref{fig_continuous} (a) and (b). In the next subsection, we show that the $\propto 1/\bar{n}_\mathrm{ex}$ scaling of the leading decay rates can be understood from a semiclassical analysis, which tells us that these eigenmodes describe a phase diffusion process.

\subsection{Continuous time crystal: semiclassical approach}

In the following, we present an effective semiclassical model that provides analytical insights into the main features of the continuous time-crystal regime. In the mean-field approximation, the effect of fluctuations is completely neglected and, therefore, information about the leading relaxation timescales of the quantum system is generally lost. In order to understand the features discussed in the previous subsection a next level of approximation is required. This can be achieved with a semiclassical approach inspired by a system-size expansion, extensively applied in nonlinear quantum optics \cite{CarmichaelBook}, e.g. in the study of dispersive optical bistability \cite{Drummond1980,Vogel1988,Vogel1989}, resonance fluorescence \cite{Drummond1978,Carmichael1980}, or in QvdP systems \cite{Lee2013,Kato2019,Kato2021} to name a few, and also in the context of cold atoms, e.g. \cite{Polkovnikov2010,Carusotto2013}. As detailed in Appendix \ref{App_semiclassical_model}, such a description can be obtained by approximating the master equation in a phase space representation by a Fokker-Planck equation \cite{CarmichaelBook,GardinerBook}.  

Focusing on the small detuning limit, i.e. $\Delta/\gamma_1\ll1$ (which implies $\eta_\mathrm{c}/\gamma_1\ll1$), intensity and phase fluctuations can be approximately decoupled (see App. \ref{App_semiclassical_model} and Ref. \cite{Kato2019}). In this limit, we consider fluctuations around the mean-field average intensity, 
\begin{equation}
\delta N=\bar{N}-\bar{n}_\mathrm{ex}
\end{equation}
which are described by the following Langevin equation \cite{CarmichaelBook,GardinerBook}:
\begin{equation}\label{Intensity_Langevin}
\begin{split} 
&\delta\dot{N}(t)=-\gamma_1\delta N(t)+\xi_{\delta N}(t),\\
&\mathbb{E}[\xi_{\delta N}(t)]=0, \quad \mathbb{E}[\xi_{\delta N}(t)\xi_{\delta N}(t')]=3\gamma_1 \bar{n}_\mathrm{ex}\delta(t-t'),
\end{split} 
\end{equation}
and the dynamics of the phase, which is described by
\begin{equation}\label{Phase_Langevin}
\begin{split} 
&\dot{\phi}(t)=-\Delta+2\eta\sin[2\phi(t)]+\xi_\phi(t),\\
&\mathbb{E}[ \xi_\phi(t)]=0, \quad \mathbb{E}[ \xi_\phi(t)\xi_\phi(t') ]=\frac{3\gamma_1}{4\bar{n}_\mathrm{ex}}\delta(t-t'),
\end{split} 
\end{equation}
where $\mathbb{E}[\dots]$ stands for an average over stochastic realizations, while $\xi_{\delta N(\phi)}(t)$ are  Gaussian white noise terms \cite{CarmichaelBook,GardinerBook}. The intensity dynamics describes a relaxation towards the mean-field intensity, while the phase dynamics retains its full non-linear form. This is commonly observed when studying the fluctuations around limit-cycle attractors, owing to the fact that while the amplitude is dynamically fixed, the phase remains free \cite{Kato2019,BenArosh2021}.

{\it Semiclassical spectrum. --}  The Langevin Eqs. (\ref{Intensity_Langevin}) and (\ref{Phase_Langevin}) can be equivalently formulated as a Fokker-Planck equation for the probability distribution $W(\delta N,\phi,t)$ (see App. \ref{App_semiclassical_model}):
\begin{equation}\label{FP_decoupled}
\begin{split}
\partial_t W(\delta N,\phi,t)=(\hat{L}_{\delta N}+\hat{L}_\phi) W(\delta N,\phi,t),
\end{split}
\end{equation}
where the expressions for the intensity Fokker-Planck operator, $\hat{L}_{\delta N}$, and the phase one, $\hat{L}_\phi$, are given in Eqs. (\ref{FP_intenisty}) and (\ref{FP_phase}), respectively. The eigenvalues of the Fokker-Planck operator characterize the leading timescales of the system, providing a semiclassical approximation to the Liouvillian spectrum. In our case, the eigenvalues are:
\begin{equation}
\tilde{\lambda}_{m,n}=\mu_m+\nu_n,    
\end{equation}
which are given as the sum of the {\it intensity eigenvalues} ($\mu_m$) and the {\it phase eigenvalues} ($\nu_n$). The intensity eigenvalues can be found analytically  (see App. \ref{App_inetnsity_eigs}):
\begin{equation}\label{intensity_eigenvalues}
\mu_m=-m\gamma_1, \quad m=0,1,2,\dots 
\end{equation}
Using first-order perturbation theory in the small parameter $1/\bar{n}_\mathrm{ex}$, the phase eigenvalues can be approximated by (see App. \ref{App_phase_eigs}):
\begin{equation}\label{phase_eigenvalues}
\nu_n=in\Omega -\frac{\gamma_1}{\bar{n}_\mathrm{ex}} c_n, \quad n=0,\pm1,\pm2,\dots   
\end{equation}
Here $c_n$ is a positive real constant that depends on the unperturbed eigenmodes and on $\Omega$. Notice that $c_0=0$ (the stationary state) while we numerically find that $c_n=n^2 c_1$, forming a ladder structure, which for $\eta=0$  can be analytically shown to be $c_1=3/8$.

{\it Band structure and phase diffusion. --} A crucial difference between the intensity and phase eigenspectrum is that the former remains gaped as the number of excitations increases, while the phase eigenspectrum becomes purely imaginary as $\bar{n}_\mathrm{ex}\to\infty$. This explains the band structure of the leading eigenmodes of the Liouvillian observed in Fig. \ref{fig_continuous} (c); the fundamental band of eigenvalues corresponds to pure phase eigenvalues, i.e. $\tilde{\lambda}_{0,n}=\nu_n$, while the subsequent higher order bands contain mixed intensity and phase eigenvalues, i.e. $\tilde{\lambda}_{m,n}$ with $m\geq1$, with a gap between bands that tends to $\gamma_1$ (see App. \ref{app_continuous}). Thus, the Liouvillian eigenmodes behind the emergence of non-decaying oscillations correspond only to the phase dynamics, such that $\lim_{\bar{n}_\mathrm{ex}\to\infty} \nu_n = i n\Omega$.

In Fig. \ref{fig_continuous} (d) we compare the real parts for the first four eigenvalues for the Liouvillian and the different approximations presented in this section: in blue circles the exact results for the Liouvillian; in red empty circles the exact results for  $\hat{L}_\phi$; while in black dashed lines the perturbative approximation [the real part of Eq. (\ref{phase_eigenvalues})]. We observe an excellent agreement between the exact and semiclassical results for sufficiently large  $\bar{n}_\mathrm{ex}$. The semiclassical approximation allows us to analyze significantly larger values of $\bar{n}_\mathrm{ex}$ than the full quantum model (see also App. \ref{App_phase_difusion}).  Therefore, Eq. (\ref{phase_eigenvalues}) provides a quantitative prediction about the lifetime of the emerging oscillations, which increases proportionally to the mean-field excitation number, and thus diverges in the infinite-excitation limit. Moreover, this allows us to identify  {\it phase diffusion} as the fluctuation process behind the finite lifetime of the oscillations for finite sizes. This is because the leading eigenmodes correspond to phase-only eigenmodes, and describe the diffusive dynamics of this degree of freedom.

\subsection{Parity symmetry breaking: full quantum model}

The structure of the Liouvillian spectrum changes qualitatively as the squeezing strength is increased above the critical point $\eta>\eta_\mathrm{c}$. This change has important consequences as the limit of large $\bar{n}_\mathrm{ex}$ is approached. In the continuous time-symmetry breaking regime all complex eigenvalues $\lambda_{j\neq 0}$ were distributed in a curved many-fold of as  in Fig. 
\ref{fig_continuous} (c). In the parity symmetry breaking regime instead, as shown in Fig. 1 of Ref. \cite{Cabot2021}, all the leading eigenvalues are real, and the characteristic band structure reported in the previous section disappears.

As the number of excitations increases, a spectral gap opens between $\lambda_j$ for $j\geq 2$ and $\lambda_{0,1}$, while simultaneously $\lambda_1\to 0$. In Ref. \cite{Cabot2021} we analyzed the consequences of the gap opening between $\lambda_{0,1}$ and $\lambda_{j\geq 2}$, in which we characterized in detail a resulting effective metastable dynamics and its relation with quantum entrainment. Here instead, we  focus on how,  for $\eta>\eta_\mathrm{c}$, the Liouvillian gap closes as $\bar{n}_\mathrm{ex}\to\infty$. In other words, we will study the limit $\lambda_1\to 0$,  where a second stationary state emerges, resulting in the parity symmetry breaking \cite{Minganti2018}.

\begin{figure}[t!]
 \centering
 \includegraphics[width=1\columnwidth]{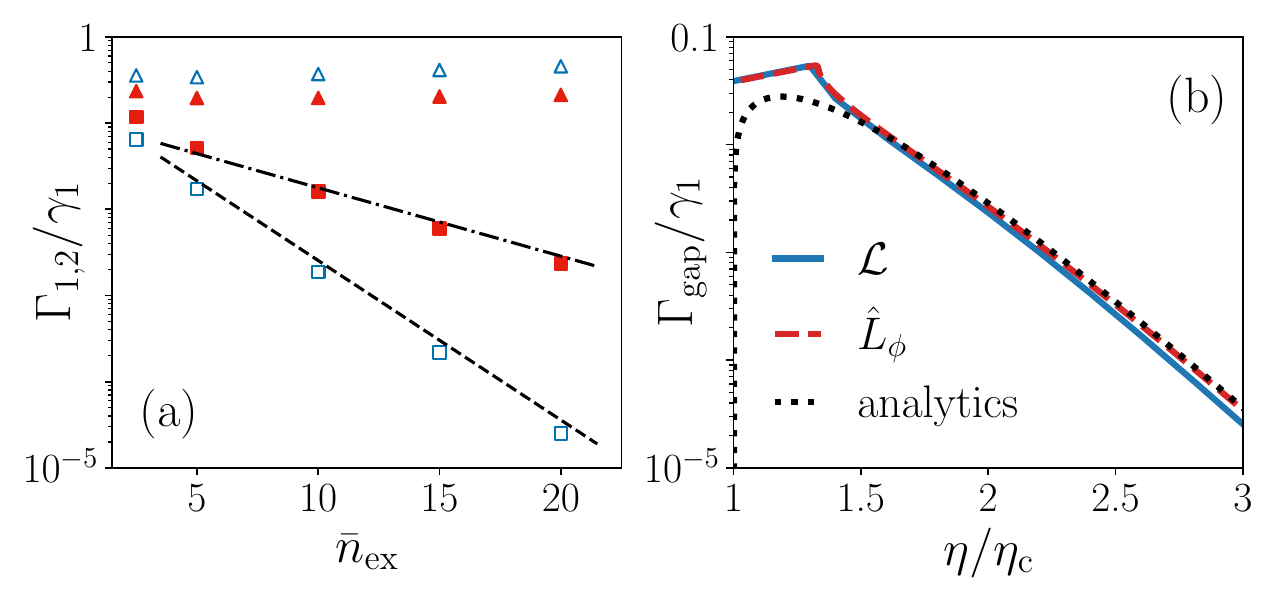}
 \caption{{\bf Signatures of parity symmetry breaking.}    (a)   $\Gamma_1/\gamma_1$ as a function of  $\bar{n}_\mathrm{ex}$ (in log scale) for $\eta/\eta_\mathrm{c}=2$ (filled red squares) and $\eta/\eta_\mathrm{c}=3$ (empty blue squares). $\Gamma_2/\gamma_1$ varying $\bar{n}_\mathrm{ex}$ for $\eta/\eta_\mathrm{c}=2$ (filled red triangles) and $\eta/\eta_\mathrm{c}=3$ (empty blue triangles). Black dashed and dashed-dotted lines correspond to the analytical result Eq. (\ref{Liouvillian_gap}) for the corresponding parameter values. (b) Blue solid line: Liouvillian gap (i.e. smallest decay rate) for $\bar{n}_\mathrm{ex}=20$. Dashed red line: smallest decay rate of the phase Fokker-Planck operator $\hat{L}_\phi$. Black dotted line: analytical approximation of the smallest decay rate [Eq. (\ref{Liouvillian_gap})]. In all panels  $\Delta/\gamma_1=0.1$.}
 \label{fig_parity}
\end{figure}

{\it Closure of the Liouvillian gap. --} In Fig. \ref{fig_parity} (a) we show the behavior of the two leading eigenvalues with $\bar{n}_\mathrm{ex}$ and for two different squeezing strengths larger than the critical one. As we can see $\lambda_2$ saturates to a finite value (triangles) while $\lambda_1$ diminishes with $\bar{n}_\mathrm{ex}$ (squares). In fact, $\lambda_1$ vanishes exponentially with the mean-field excitation number. This is confirmed when comparing the squares with the black dashed lines, which correspond to the analytical result presented below [Eq. (\ref{Liouvillian_gap})]  and which predicts that the Liouvillian gap vanishes exponentially with $\bar{n}_\mathrm{ex}$. We find excellent agreement between the two up to a slight overestimation of the gap.
This provides strong evidence for the gap closure in the infinite-excitation limit. Alternatively, the Liouvillian gap can also be made to close by increasing the squeezing strength (which also increases the excitation number). This is shown in Fig.  \ref{fig_parity} (b) (blue solid line) and, as the black dashed line shows, it is also predicted by our analytical approach.

{\it Symmetry broken states. --} The closure of the Liouvillian gap is the signature of a dissipative phase transition \cite{Kessler2012,Minganti2018} associated with a symmetry breaking. Indeed, here, the closing eigenmode, $\hat{r}_1$, belongs to a different symmetry sector with respect to
$\hat{\rho}_\mathrm{ss}$ implying the breakdown of parity symmetry in this regime \cite{Minganti2018}. Recall that as a consequence of the parity symmetry of the Liouvillian, its eigenmodes transform as \cite{Albert2014,Minganti2018}:
\begin{equation}
\mathcal{Z}_2\hat{r}_j=z_j\hat{r}_j    
\end{equation}
where $z_j=\pm1$ is the parity eigenvalue. The stationary state belongs to the symmetric sector \cite{Albert2014,Minganti2018}, i.e. $z_0=1$, while numerical analysis reveals that $\hat{r}_1$ is antisymmetric, i.e. $z_1=-1$. In these conditions, and in the limit $\eta>\eta_\mathrm{c}$, $\bar{n}_\mathrm{ex}\to\infty$ in which the gap closes, we can decompose $\hat{r}_1$ and $\hat{\rho}_\mathrm{ss}$ in terms of {\it symmetry broken stationary states} (see Ref. \cite{Minganti2018} and App. \ref{app_parity}):
\begin{equation}
\hat{\rho}_\pm=\hat{\rho}_\mathrm{ss}\pm\hat{r}_1.    
\end{equation}
These states satisfy: $\mathcal{L}\hat{\rho}_\pm=0$ and $\mathcal{Z}_2\hat{\rho}_\pm=\hat{\rho}_\mp$, hence being stationary and  parity symmetry broken. Indeed, they are mapped one to each other through a parity transformation, analogously to the mean-field solutions [Eq. (\ref{MF_solution})], while the observables computed on these states tend to the corresponding mean-field stable solutions (see App. \ref{app_parity}). Therefore, the emergence of these symmetry-broken states through the Liouvillian gap closure signals the emergence of the mean-field bistable regime as the infinite-excitation limit is approached.

\subsection{Parity symmetry breaking: semiclassical approach}

\begin{figure}[t!]
\includegraphics[width=0.9\linewidth]{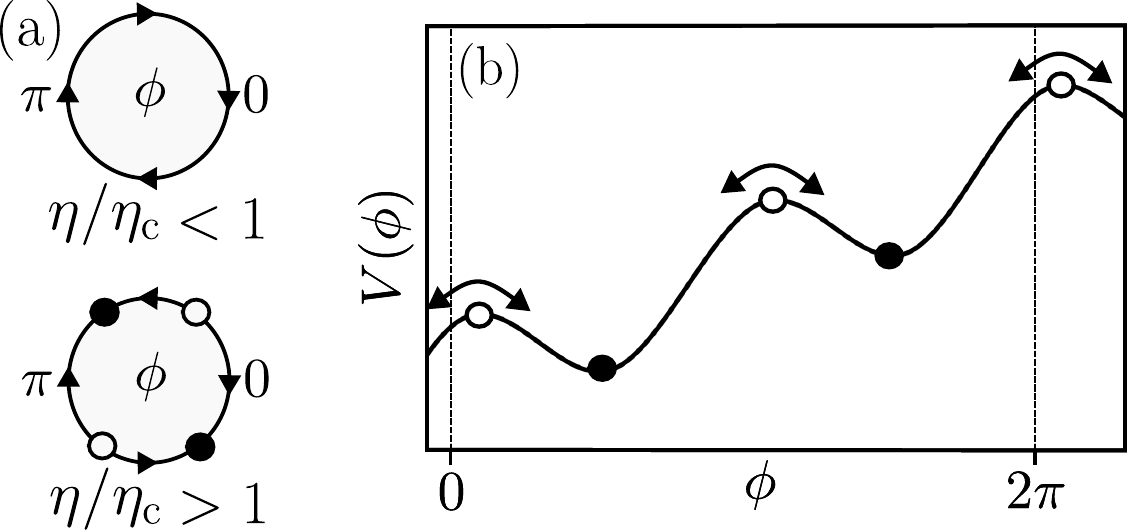}\\
\caption{{\bf Quantum activation process}. (a) Flow diagram of the phase degree of freedom in the limit-cycle (top) or the bistable (bottom) regime. Stable fixed points are represented with full circles and unstable ones with empty circles.  (b) Tilted washboard potential $V(\phi)$   [Eq. (\ref{Potential_Phase})] in the bistable regime. In the the interval $\phi\in[0,2\pi)$ there are two minima (corresponding to the stable fixed points) and two maxima (corresponding to the unstable ones). The activation process can be thought of as follows: supposing the system starts in the left stable point it can transition to the other state jumping over the right barrier with a rate $\Gamma_\rightarrow$ [Eq. (\ref{gamma_right})]  or over the left one (since the potential is periodic) with a rate  $\Gamma_\leftarrow$ [Eq. (\ref{gamma_left})]. This is represented by the arrows.  In this case, jumping to the right is exponentially suppressed with respect to jumping to the left as the potential barrier is larger, and hence the process is dominated by the latter.}\label{fig_potential}
\end{figure}

In the presence of bistability and noise, we can generically expect two fluctuation modalities: (i) small fluctuations around the fixed points and (ii) noise-induced transitions between the two fixed points \cite{GardinerBook}. In potential systems and in the small noise limit, noise-induced transitions occur at a rate that decreases exponentially with the ratio of the potential barrier over the noise intensity \cite{GardinerBook}.  Thus, process (ii) can be expected to govern the longest timescale dynamics in the system. Such a fluctuation process is ruled by  the phase equation  (\ref{Phase_Langevin}) {whose  deterministic part  corresponds to the derivative of the phase potential
\begin{equation}\label{Potential_Phase}
V(\phi)=\Delta\phi+\eta\cos(2\phi).
\end{equation}
For $\eta>\eta_\mathrm{c}$ this displays two minima in the range $\phi\in[0,2\pi)$ corresponding to the bistable fixed points (see Fig. \ref{fig_potential}). Thus, the noise term in Eq. (\ref{Phase_Langevin}) can induce transitions between these two solutions, as illustrated in Fig. \ref{fig_potential} (b).
In these conditions, we can obtain an approximate expression for the rate of jumps over the potential barriers (see App. \ref{App_quantum_activation}) \cite{GardinerBook}:
\small
\begin{equation}\label{Liouvillian_gap}
\Gamma_\mathrm{gap}=C \exp\bigg[-\frac{8\bar{n}_\mathrm{ex}}{3\gamma_1}\bigg(\sqrt{4{\eta}^2-{\Delta}^2}+{\Delta}\sin^{-1}\frac{|{\Delta}|}{2{\eta}}-\frac{|{\Delta}| \pi}{2}\bigg)\bigg] ,  
\end{equation}
\normalsize
with
\begin{equation}
C=\frac{2}{\pi}\sqrt{4{\eta}^2-{\Delta}^2}    ,
\end{equation}
which we can use as an approximation for the spectral gap of $\hat{L}_\phi$ and the Liouvillian. Indeed in Fig. \ref{fig_parity} (b) we compare the Liouvillian gap (blue solid line) with the gap of $\hat{L}_\phi$ (red dashed line) and this analytical result (black dotted line). From this figure we observe that  Eq. (\ref{Liouvillian_gap}) is very accurate away from the transition, only slightly overestimating the Liouvillian gap.  Finally, we note that just at the transition point ($\eta=\eta_c$) we find the Liouvillian gap to display a different scaling law, being proportional to $\lambda_1\propto \bar{n}_\mathrm{ex}^{-0.37}$. Interestingly, power law scalings have been also observed at the transition point of other nonequilibrium systems \cite{Buonaiuto2021,Wang2021}.

{\it Quantum activation process. --} The excellent agreement between Eq. (\ref{Liouvillian_gap}) and of the Liouvillian gap leads us to conclude that the dominant fluctuation process in the bistable regime is that  of {\it quantum activation} \cite{Dykman2006,Dykman2007,Dykman2018,Gosner2020}. This is an incoherent phenomenon in which the system ``jumps'' over potential barriers because of  the accumulated effect of quantum fluctuations. The fact that the phase equation captures the correct eigenvalue behavior implies that this equation accurately describes the preferred direction of jumping in phase space between the two stable solutions, i.e., the phase space path through which it is easier to transition from one to the other \cite{GardinerBook}. 

\section{Spontaneous symmetry breaking in the laboratory frame}\label{Sec_discrete}

In this section we address the dynamics of the full quantum model in the laboratory frame and analyze the spontaneous breaking of the discrete time-translation symmetry. This occurs when stable solutions emerge that do not respect the periodicity of the Liouvillian, $\mathcal{L}_\mathrm{L}(t+T)=\mathcal{L}_\mathrm{L}(t)$. In the case of a discrete time crystal, this is  signaled by observables that display non-decaying oscillations whose period is a {\it multiple} of $T$:
\begin{equation}\label{timecrystals_Discrete}
f_\mathrm{L}(\tau+nT)=f_\mathrm{L}(\tau), \quad n>1. 
\end{equation}
as shown in, e.g., Refs \cite{Gong2018,Wang2018,Dykman2018,Gambetta2019,Lazarides2020,Riera2020,Zhu2019,Chinzei2020,Kessler2021,Tuquero2022,Chitra2015}. In this frame. quasiperiodic solutions can also emerge, in which the dynamics contains components oscillating at incommensurate frequencies. This leads to the emergence of an incommensurate time crystal \cite{Cosme2019,Chinzei2020,Homann2020,Skulte2021,Nie2023,Cosme2023}.

{\it Stroboscopic dynamics in the laboratory frame. --} For our purposes, it is sufficient to consider the state in the laboratory frame at stroboscopic times, i.e. $\hat{\rho}_\mathrm{L}(t=nT)$ with $n$ a positive integer. In this way, we can understand the dynamics in the laboratory frame from the results we already have for the rotating frame. This is because at multiples of the Hamiltonian period ($t=nT$)  the unitary transformation (\ref{lab_to_rot}) that links both frames is equivalent to a set of $n$ concatenated parity transformations:
\begin{equation}
\hat{\rho}_\mathrm{L}(nT)=\hat{U}_{nT} \hat{\rho}(nT)\hat{U}_{nT}^\dagger\\
=e^{-in\pi\hat{a}^\dagger\hat{a}}\hat{\rho}(nT)e^{in\pi\hat{a}^\dagger\hat{a}}.
\end{equation}
Owing to the fact that for finite $\bar{n}_\mathrm{ex}$ the eigenmodes have a well-defined parity symmetry, we can also decompose the {\it stroboscopic laboratory frame dynamics} in their terms:
\begin{equation}\label{strob_times}
\hat{\rho}_\mathrm{L}(nT)=\hat{\rho}_\mathrm{ss}+\sum_{j\geq1}(z_j)^n\,\text{Tr}[\hat{l}^\dagger_j\hat{\rho}(0)]\hat{r}_j e^{\lambda_j nT}.   
\end{equation}
Depending on the parity of the eigenmodes ($z_j=\pm1$) the additional oscillating phase $(-1)^n$ is displayed. Eq. (\ref{strob_times}) tells us that the dynamics in the laboratory frame can be fully analyzed if we know the eigenvalues $\lambda_j$,  their behavior when approaching the limit $\bar{n}_\mathrm{ex}\to\infty$, and the parity $z_j$ of their respective eigenmodes. 

\subsection{Emergence of an incommensurate time crystal for $\eta<\eta_\mathrm{c}$}

For $\eta<\eta_\mathrm{c}$ and approaching the limit $\bar{n}_\mathrm{ex}\to \infty$, the eigenvalues of the fundamental band display a vanishing decay rate and an imaginary part that tends to $\Omega$ and multiples of it (see Fig. \ref{fig_continuous} and App. \ref{app_continuous}). For long times and in the infinite-excitation limit only this fundamental band of eigenmodes contributes to the dynamics, and we can write the state of the system as: 
\begin{equation}\label{strob_times_quasicrystal}
\hat{\rho}_\mathrm{L}(nT)\approx\hat{\rho}_\mathrm{ss}+\sum_{j\in \{\Gamma_j=0\}}(z_j)^n\,\text{Tr}[\hat{l}^\dagger_j\hat{\rho}(0)]\hat{r}_j e^{in \Omega_j T},   
\end{equation}
with 
\begin{equation}
\Omega_j\in\{m\Omega, \quad m=\pm1,\pm2,\pm 3,\dots\}.
\end{equation}
In the infinite-excitation limit the approximation sign accounts for neglecting the remaining modes of the higher-order bands at large times. In Eq. (\ref{strob_times_quasicrystal}) we use the notation $\Omega_j$ to denote that the frequency can be $\Omega$ or a multiple of it. In all cases, we can rewrite the time-dependent factor as:
$e^{i\Omega_j nT}=e^{i n\pi \frac{\Omega_j}{\omega_\mathrm{s}}}.$
The frequency $\Omega$ varies continuously with the system parameters [see Eq. (\ref{freq_LC})] and thus, without fine-tuning, it is generally incommensurate with $\omega_\mathrm{s}$. This means that the non-decaying time-dependent solution described by Eq. (\ref{strob_times_quasicrystal}) is incommensurate with the driving period. Moroever, since Eq. (\ref{strob_times_quasicrystal}) contains components both at $\Omega$ and its harmonics, the emergent solution is generally {\it quasiperiodic}. Therefore, for $\eta<\eta_\mathrm{c}$ and in the laboratory frame the system displays an incommensurate time crystal.

\subsection{Emergence of a discrete time crystal for $\eta>\eta_\mathrm{c}$}

\begin{figure}[t!]
 \centering
 \includegraphics[width=1\columnwidth]{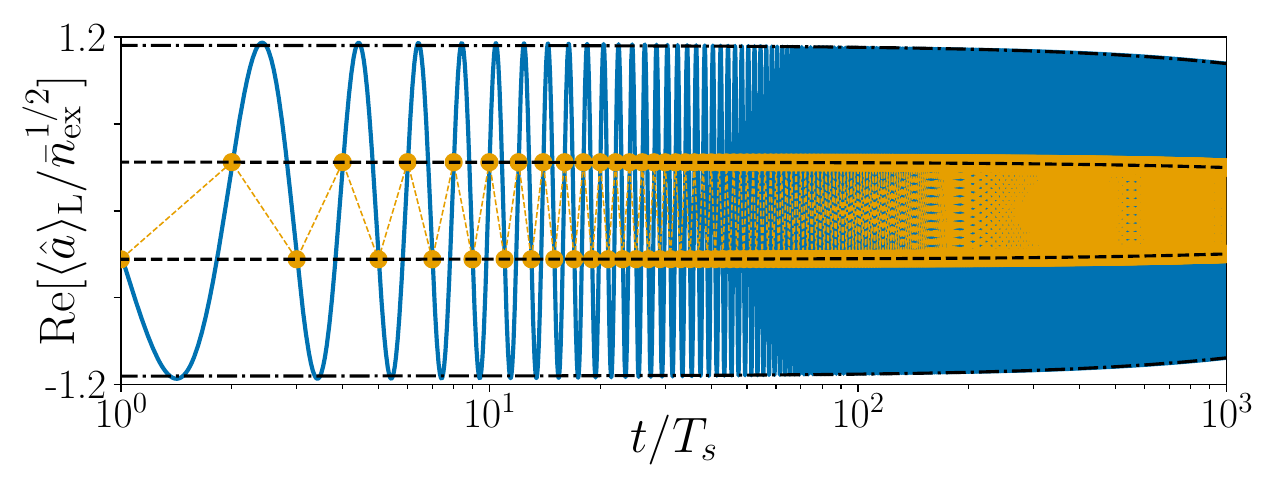}
 \caption{{\bf Signatures of discrete time-translation symmetry breaking.}   Amplitude dynamics in the laboratory frame for $\omega_\mathrm{s}/\gamma_1=20\pi$, $\eta/\eta_\mathrm{c}=2$, $\Delta/\gamma_1=0.1$, $\bar{n}_\mathrm{ex}=20$ and a coherent state of amplitude $\alpha_+$ as initial condition. The blue solid line corresponds to the full-time evolution; orange line points correspond to the stroboscopic time evolution; black dashed and dashed-dotted lines correspond to exponential decaying envelopes with decay rate $\Gamma_\mathrm{gap}/\gamma_1$.}
 \label{fig_discrete}
\end{figure}

Next, we show that the parity-breaking DPT occurring in the rotating frame for $\eta >\eta_\mathrm{c}$ implies discrete time-translation symmetry breaking in the laboratory frame. This can be demonstrated by use of Eq. (\ref{strob_times}) and the results presented in Sec. \ref{Sec_continuous}. As the Liouvillian gap closes in the infinite-excitation limit and $\hat{r}_1$ is parity antisymmetric ($z_1=-1$), then for $nT\gg\Gamma_2^{-1}$ and $\bar{n}_\mathrm{ex}\to\infty$ the state of the system reads
\begin{equation}\label{timecrystals_DTC}
\hat{\rho}_\mathrm{L}(nT)\approx \hat{\rho}_\mathrm{ss}+(-1)^n\,\text{Tr}[\hat{l}^\dagger_1\hat{\rho}(0)]\hat{r}_1.   
\end{equation} 
In the infinite-excitation limit the approximation sign accounts for neglecting the remaining modes ($j>1$) at large times. Eq. (\ref{timecrystals_DTC}) clearly signals the breakdown of the discrete time-translation symmetry, as the state of the system shows a non-decaying period-doubled dynamics. Parity antisymmetric observables are capable of resolving this period doubling, and thus, they can be used as an order parameter of the type given in Eq.  (\ref{timecrystals_Discrete}). This is illustrated in Fig. \ref{fig_discrete}, in which we plot $\langle\hat{a}(t)\rangle_\mathrm{L}$ solving the master equation in the laboratory frame both at stroboscopic times (yellow squares) and continuous time (blue solid lines). We can see that the stroboscopic dynamics alternate between two values that depend on the initial condition, while the full-time evolution depicts the complete harmonic oscillation of frequency $\omega_\mathrm{s}$. Since the considered $\bar{n}_\mathrm{ex}$ is finite, the symmetry broken state has a finite lifetime and thus this period-doubled dynamics eventually decays out on a timescale $\Gamma_\mathrm{gap}^{-1}$. This decaying envelope is shown in black-dashed lines.

\section{Exceptional point as the bifurcation point}\label{Sec_ep}

In the previous sections, we have shown that the Liouvillian spectrum displays qualitatively different features in the different regimes as the infinite-excitation limit is approached. In the rotating frame, the continuous time-crystal is signaled by a whole band of eigenmodes displaying a vanishing decay rate for $\eta<\eta_c$, while  for $\eta>\eta_c$  the Liouvillian gap closes and the parity symmetry is spontaneously broken. Here we address how the spectrum changes from one regime to the other for \textit{any} value of $\bar{n}_\mathrm{ex}$.   Interestingly, we show that a spectral singularity, an exceptional point, mediates in between both regimes and plays an analogous role to the bifurcation point. This singular behavior {\it for finite excitation number} is in stark contrast with the smooth behavior of the observables in the steady state in the quantum regime (see e.g. Fig. \ref{fig_diagram}). 

\begin{figure}[t!]
\includegraphics[width=1\linewidth]{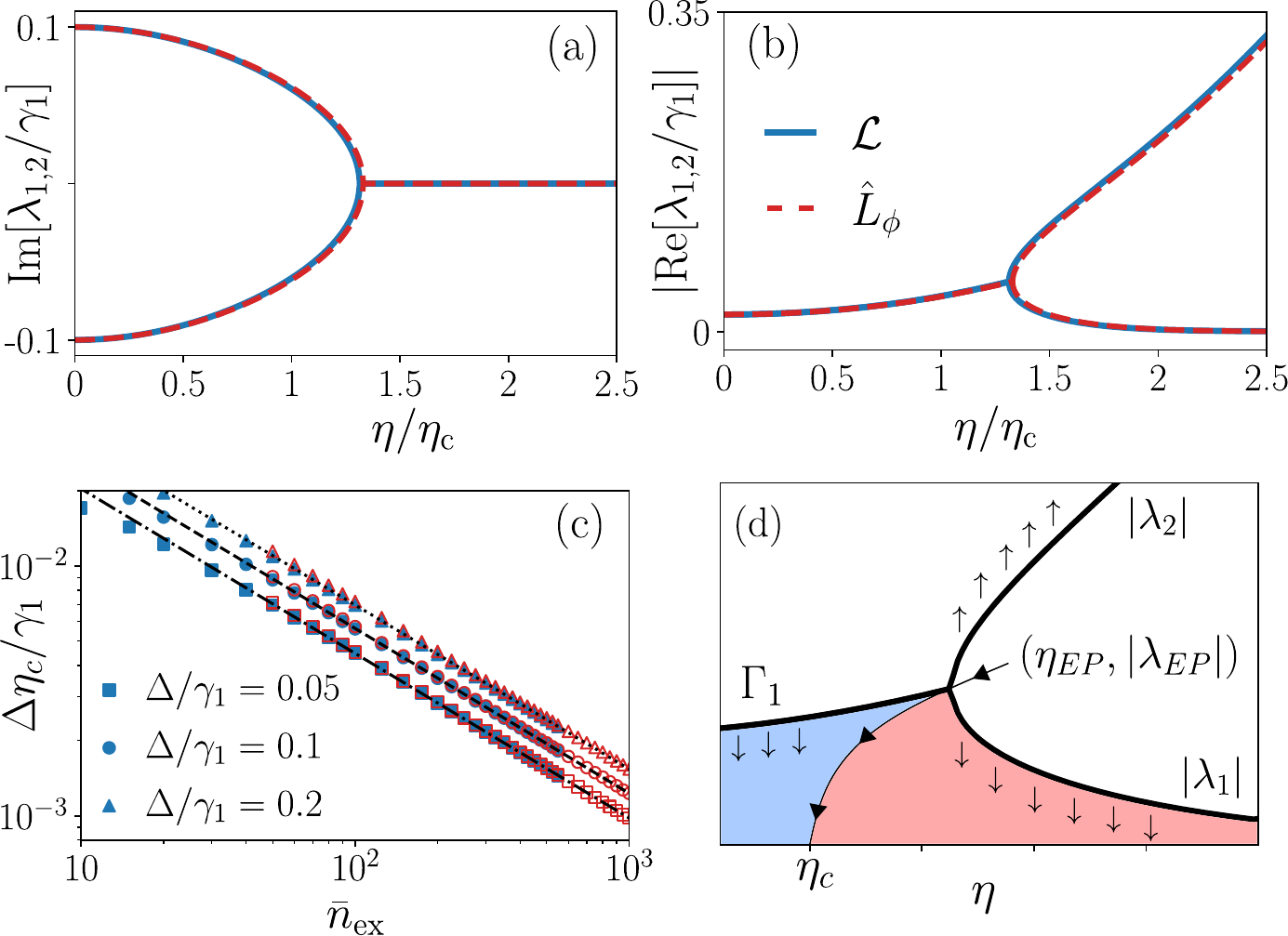}\\
\caption{{\bf Exceptional point and emerging bifurcation.} Imaginary (a) and real (b) parts of the two dominant eigenvalues varying the squeezing strength across the transition for $\bar{n}_\mathrm{ex}=20$.  The blue solid lines correspond to the eigenvalues of the Liouvillian, while the red dashed lines correspond to those of the phase Fokker-Planck operator. (c) $\Delta\eta_\mathrm{c}=(\eta_{EP}-\eta_\mathrm{c})$ varying $\bar{n}_\mathrm{ex}$ and for different values of $\Delta/\gamma_1$. Blue full markers correspond to the results obtained from the Liouvillian, while red empty markers correspond to those obtained from the phase Fokker-Planck operator. Black-dashed lines correspond to the fit  $\Delta\eta_\mathrm{c}/\gamma_1\propto \bar{n}_\mathrm{ex}^{-\beta}$ for for $\bar{n}_\mathrm{ex}\in[100,550]$ for the Liouvillian. The exponents read $\beta=[0.659,0.660,0.656]$ for $\Delta/\gamma_1=[0.05,0.1,0.2]$ respectively, and where the last decimal is within the standard deviation of the fit. The corresponding fits for the semiclassical results for $\bar{n}_\mathrm{ex}\in[500,10^4]$ yield the exponents  $\beta=[0.670,0.671,0.671]$. (d) Cartoon of the emerging bifurcation diagram and the behavior of the leading eigenvalues of the Liouvillian as the squeezing strength is varied. The black arrows depict the qualitative behavior of the leading decay rates (in the continuous time-crystal regime), the two dominant eigenvalues (in the parity symmetry breaking regime) and the EP, as the infinite-excitation limit is approached.}\label{fig_EP}
\end{figure}

{\it Exceptional point (EP). --} In Fig. \ref{fig_EP} we show the behavior of the imaginary (a) and real (b) parts of the two leading Liouvillian eigenvalues (blue solid lines) varying the squeezing strength, $\eta$, across the transition point, $\eta_\mathrm{c}$. For small $\eta/\eta_\mathrm{c}$, the two eigenvalues are complex conjugate, while at $\eta=\eta_\mathrm{EP}$ they collide in the complex plane and they become real for larger squeezing. The point of degeneracy corresponds to an EP, in which both the eigenvalues and eigenvectors coalesce and the Liovillian becomes singular \cite{Heiss2012,Minganti2019}. The presence of this EP was first reported in Ref. \cite{Cabot2021}, and lies in between the parameter regimes in which the eigenvalues behave in a qualitatively different fashion. This EP is also captured by the semiclassical phase model [Eq. (\ref{Phase_Langevin})], as we show in red dashed lines. Indeed, we observe an excellent agreement between the two leading eigenvalues of $\mathcal{L}$ and of $\hat{L}_{\phi}$, despite the EP in the latter seems to occur at a slightly larger squeezing strength. The EP can also be observed in Fig. \ref{fig_parity} (b) as the tipping point in which the Liouvillian gap changes its behavior. We recall that the EP also manifests in the laboratory frame, as the dynamics at stroboscopic times [Eq. (\ref{strob_times})] depends on the spectrum of the Liouvillian in the rotating frame.

{\it EP and bifurcation point. --} From  Fig. \ref{fig_EP} (a), (b), the question arises whether $\eta_\mathrm{EP}$ and $\eta_\mathrm{c}$ are related. As a matter of fact, we observe that $\eta_\mathrm{EP}$  depends on the other parameters of the system, i.e., the detuning and the mean-field excitation number. We observe that $\eta_\mathrm{EP}>\eta_\mathrm{c}$ for finite $\bar{n}_\mathrm{ex}$ while their difference vanishes as the infinite-excitation limit is approached. This is shown in Fig. \ref{fig_EP} (c), where we consider three different values of the detuning, and we analyze the difference between the bifurcation point and the EP, i.e. $\Delta\eta_\mathrm{c}=\eta_\mathrm{EP}-\eta_\mathrm{c}$  for the Liouvillian (blue full markers) and for the phase Fokker-Planck operator (red empty markers), the latter allowing us to explore significantly larger values of $\bar{n}_\mathrm{ex}$. In all cases we find that the following scaling law  accurately captures the asymptotic value of this difference:
\begin{equation}\label{EP_exponent}
\Delta\eta_\mathrm{c}/\gamma_1\propto \bar{n}_\mathrm{ex}^{-\beta}.
\end{equation}
The black dashed lines in Fig. \ref{fig_EP} (c) correspond to the fit to the points with the largest $\bar{n}_\mathrm{ex}$ for the Liouvillian, in which we find a similar exponent for the three detunings, namely  $\beta\sim 0.66$. Considering the semiclassical results (red empty markers), we see that they slightly overestimate the Liouvillian ones, however, they are still accurately fitted by the same scaling law [Eq. (\ref{EP_exponent})]. In this case, we have performed the same fit for much larger excitation numbers (up to $\bar{n}_\mathrm{ex}=10^4$) obtaining similar exponents $\beta\sim 0.67$. This strengthens the conclusion that as the infinite-excitation limit is approached the EP approaches the bifurcation point as a power law in $\bar{n}_\mathrm{ex}$, i.e. $\lim_{\bar{n}_\mathrm{ex}\to \infty}\eta_\mathrm{EP}=\eta_\mathrm{c}$. Therefore, the EP provides a spectral signature of the emergent nonequilibrium transition.

\section{Summary and conclusions}\label{sec_conclusions}

In this work, we have analyzed a dissipative phase transition that occurs in the squeezed QvdP oscillator. This transition separates two distinct regimes in which different symmetries are spontaneously broken. In the laboratory frame, the system breaks the discrete time-translation symmetry in two different ways. When the squeezing strength is smaller than the critical one, an incommensurate time crystal emerges, with its characteristic non-decaying quasiperiodic response. When the squeezing strength is larger, a discrete time crystal emerges. In the rotating frame, the former manifests as a continuous time crystal, while the latter manifests as a parity symmetry broken phase (see Table \ref{tab:table1}).

Addressing the Liouvillian spectral signatures of both phases, we have shown them to be accurately described by a semiclassical fluctuation model in which the dominant fluctuation modes correspond to a phase degree of freedom. This has allowed us to identify the main fluctuation processes in each regime as well as to obtain accurate expressions for the dominant Liouvillian eigenvalues. We have shown that {\it phase diffusion} is behind the finite lifetime of the oscillations in the continuous time crystal (or incommensurate time crystal in the laboratory frame), while {\it quantum activation} \cite{Dykman2006,Dykman2007,Dykman2018,Gosner2020} accounts for the dominant lifetime in the parity breaking regime (discrete time crystal in the laboratory frame). These processes display qualitatively different spectral signatures as the asymptotic scaling of the dominant lifetimes:  for phase diffusion the dominant lifetime scales linearly with the mean-field excitation number ($\bar{n}_\mathrm{ex}$), while for quantum activation regime, it scales exponentially with it. 

The peculiar nonequilibrium transition we have reported manifests as a spectral singularity even for finite excitation numbers. For finite $\bar{n}_\mathrm{ex}$, the point where the eigenspectrum changes qualitatively is an exceptional point (EP) occurring at $\eta_\mathrm{EP}$, where the leading eigenvalues  $\lambda_{1,2}$ become real. For smaller squeezing strengths, the spectrum develops the band structure characteristic of phase diffusion, while for larger ones, the dominant eigenvalue vanishes and the rest remains gaped as $\bar{n}_\mathrm{ex}\to\infty$. The squeezing strength at which the EP occurs approaches the classical bifurcation point as $\bar{n}_\mathrm{ex}\to\infty$ following a power law in the mean-field excitation number, as we have shown for the Liouvillian and semiclassical spectra. These results are graphically summarized in Fig. \ref{fig_EP} (d), which shows the qualitative behavior of the leading eigenvalues and the EP with $\bar{n}_\mathrm{ex}\to\infty$. This correspondence between EP and a bifurcation point has been also observed in Ref. \cite{Minganti2018} for a different DPT, in which a different exponent $\beta$ has been identified. It remains an open question the discovery of further examples of this correspondence, and whether the exponent $\beta$ can be used to classify different families of nonequilibrium transitions.

Finally, we discuss our results in connection with the phenomenon of entrainment, i.e. synchronization to an external signal \cite{PikovskyBook}. In this system the presence or absence of entrainment is related to the time-crystal order; in the incommensurate time crystal regime there is no entrainment, while in the discrete one, there is entrainment. This can be understood by recalling that in an entrained regime, a system oscillates at  {\it the same} or at a {\it multiple/fraction} of the frequency of a forcing, which means that it either respects the time-translation symmetry imposed by the forcing (as in the driven QvdP oscillator \cite{Lee2013,Walter2014}), or it breaks it discretely, as in the present case. On the other hand, in the absence of entrainment, the system oscillates at a frequency that depends on the internal parameters and it is incommensurate with that of the forcing \cite{BelanovBook,PikovskyBook}, leading to the possible emergence of an inconmensurate time crystal. This connection between entrainment and time-crystalline order differs from that of other synchronization phenomena reported in the literature, as in Ref. \cite{Tucker2018} in which the regime of mutual synchronization does indeed correspond to a continuous time crystal. However, this distinction can be attributed to the underlying synchronization scenarios being considered. In the case of entrainment, an external frequency is imposed. Conversely, mutual synchronization occurs when a shared or common frequency emerges spontaneously among the interacting components.

\section{Acknowledgements}

We thank R. Fazio and D. Gomila for interesting discussions. We acknowledge the use of Qutip python library \cite{Qutip1,Qutip2} for performing some of the numerical calculations of this work. AC is grateful for financing from the Baden-Württemberg Stiftung through Project No. BWSTISF2019-23 and from the Deutsche Forschungsgemeinschaft
(DFG, German Research Foundation) through the
Walter Benjamin programme, Grant No. 51984724. GLG is funded by the Spanish  Ministerio de Educaci\'on y Formaci\'on Profesional/Ministerio de Universidades and co-funded by the University of the Balearic Islands through the Beatriz Galindo program (BG20/00085).
 We acknowledge the Spanish State Research Agency, through the Mar\'ia de Maeztu project CEX2021-001164-M funded by the MCIN/AEI/10.13039/501100011033, through the QUARESC project (PID2019-109094GB-C21 and -C22/ AEI / 10.13039/501100011033) and through the COQUSY project PID2022-140506NB-C21 and -C22 funded by MCIN/AEI/10.13039/501100011033, MINECO through the QUANTUM SPAIN project, and EU through the RTRP - NextGenerationEU within the framework of the Digital Spain 2025 Agenda.

\appendix

\section{Derivation of the semiclassical model and eigenvalue problem}\label{App_semiclassical_model}

In this appendix, we discuss how to obtain the semiclassical fluctuation model of Eqs. (\ref{Intensity_Langevin})-(\ref{FP_decoupled}). We begin in Sec. \ref{sec_A_1}  by rewriting the master equation in the Wigner phase space representation and we perform the semiclassical approximation of truncating higher-order partial derivatives. Afterwards, in Sec. \ref{sec_A_2}, we discuss the approximation of decoupling phase and intensity fluctuations. Finally, in Sec. \ref{sec_A_3} we overview the formulation of the eigenvalue problem for Fokker-Planck equations.

\subsection{Truncated Wigner approximation and change of variables}\label{sec_A_1}

The starting point is the exact rewriting of the master equation in the rotating frame as a partial differential equation for the Wigner distribution  \cite{CarmichaelBook}:
\small
\begin{equation}\label{eq_wigner}
\begin{split}
\partial_t W(\alpha,\alpha^*,t)=\hat{L}_\mathrm{ME} W(\alpha,\alpha^*,t)
\end{split}
\end{equation}
\normalsize
with the differential operator defined as \cite{Lee2013}
\small
\begin{equation}\label{Wigner_ME}
\begin{split}
\hat{L}_\mathrm{ME}=&\partial_\alpha \alpha\big[\gamma_2(|\alpha|^2-1)-\frac{\gamma_1}{2} \big]
+(2\eta\alpha^*+i\Delta\alpha)\partial_\alpha\\
&+\frac{\partial_\alpha\partial_{\alpha^*}}{2}\big[\frac{\gamma_1}{2}+\gamma_2(2|\alpha|^2-1)\big]+\frac{\gamma_2}{4}\partial_\alpha^2\partial_{\alpha^*}\alpha+\mathrm{c.c.}
\end{split}
\end{equation}
\normalsize
The semiclassical approximation consists in neglecting third-order partial derivatives while approximating the diffusion coefficient by the mean-field amplitude for $\eta=0$ \cite{Lee2013}, i.e. $|\alpha|^2\approx\bar{n}_\mathrm{ex}$. This approximation becomes more accurate as the limit $\bar{n}_\mathrm{ex}\to\infty$ ($\gamma_2/\gamma_1\to 0$) is approached \cite{Lee2013}, while the squeezing strength and the detuning need to be small compared with the amplitude of the limit cycle such that its amplitude and shape are not significantly altered. Notice that this approximation does not correspond to a rigorous system-size expansion \cite{CarmichaelBook}. 

Then, the {\it semiclassical} equation for the Wigner distribution reads
\begin{equation}\label{eq_FP1}
\partial_t W(\alpha,\alpha^*,t)=\hat{L}_{\text{FP}} W(\alpha,\alpha^*,t)
\end{equation}
with
%
%\small
\begin{equation}
\begin{split}
\hat{L}_{\text{FP}}=&\partial_\alpha \alpha\big[\gamma_2|\alpha|^2-\frac{\gamma_1}{2} \big]+(2\eta\alpha^*+i\Delta\alpha)\partial_\alpha\\
&+\frac{3\gamma_1}{4}\partial_\alpha\partial_{\alpha^*} +\mathrm{c.c.}  
\end{split}
\end{equation}
%\normalsize
%
Essentially, we have replaced the original quantum dynamics as described by the general partial differential equation (\ref{eq_wigner}) by a Fokker-Planck equation for the Wigner distribution \cite{CarmichaelBook}. From now on we will regard  $W(\alpha,\alpha^*,t)$ as a classical probability distribution and $\alpha$ ($\alpha^*$) as classical variables, which allows us to fully analyze Eq. (\ref{eq_FP1}) using  classical stochastic methods \cite{GardinerBook,RiskenBook}. 

From now on and in the following appendices, we work with the  dimensionless parameters:
\begin{equation}
\tau=\gamma_1 t,\quad \tilde{\Delta}=\Delta/\gamma_1, \quad \tilde{\eta}=\eta/\gamma_1.
\end{equation}
Moreover, we make a change of variables in the Fokker-Planck equation to rewrite it in terms of the phase and intensity: $\alpha=\sqrt{N}e^{i\phi}$, where $\phi$ and $N$ are the stochastic counterparts of the mean-field phase and intensity, respectively. With the use of these dimensionless parameters and the change of variables, the semiclassical model reads:
\begin{equation}
\partial_\tau W(N,\phi,\tau)=\hat{L}_{\text{FP}}W(N,\phi,\tau)
\end{equation}
with
\small
\begin{equation}\label{eq_FP2}
\begin{split}
\hat{L}_{\text{FP}}=&-\partial_N\big[1-\frac{N}{\bar{n}_\mathrm{ex}} -4\tilde{\eta}\cos(2\phi)\big]N
+\partial_\phi\big[\tilde{\Delta}-2\tilde{\eta}\sin(2\phi)\big]\\
&+\frac{3}{2}\partial_N N\partial_N+\frac{3}{8N}\partial_\phi^2.
\end{split}
\end{equation}
\normalsize
We observe that, with these new variables, noise terms become multiplicative, i.e. the diffusion terms depend on $N$. We now comment on some important aspects of Eq. (\ref{eq_FP2}): (i) the noiseless intensity dynamics depends on the phase; (ii) the noiseless phase dynamics is independent of the intensity, but it is nonlinear; (iii) the diffusion terms are exactly the same as for the case $\eta/\gamma_1=0$, and hence the strength of phase fluctuations depends inversely on the squared amplitude (intensity) of the limit-cycle, i.e. $N^{-1}$. In the following subsection, we perform some further approximations on Eq. (\ref{eq_FP2}) based on the dynamics of the system, which will allow us to obtain analytical results for the dominant timescales of the system.

\subsection{Decoupling of intensity and phase fluctuations.}\label{sec_A_2}

In the limit of small detuning and squeezing strength, i.e. $|\tilde{\Delta}|,\tilde{\eta}\ll1$, the contribution of the phase term in the {\it intensity dynamics} is small, and can be neglected in a first approximation [this accounts to drop the term proportional to $\eta$ in Eq. (\ref{MF_eq_polar})]. Recall that since the bifurcation diagram only depends on the relation between $\Delta$ and $\eta$, the considered limit does not preclude the observation of the bifurcation, as $\eta_\mathrm{c}=|\Delta|/2$ can still be small compared to $\gamma_1$. Physically, what this approximation exploits is that in this limit the amplitude of the limit cycle is large compared to the squeezing strength, and thus, it is essentially the same as that of the undriven system. Equivalently, the phase-space representation of the cycle is essentially circular, with only small elliptical deformations due to the non-zero squeezing \cite{Kato2019}. In the same way, in the bistable regime, it exploits the fact that in this limit the fixed points display an amplitude that is essentially $\sqrt{\bar{n}_\mathrm{ex}}$ plus small corrections [see Eq. (\ref{MF_solution})]. 

Therefore, in the small detuning and squeezing strength limit we approximate Eq. (\ref{eq_FP2}) by:
\begin{equation}\label{eq_FP3}
\begin{split}
\hat{L}_{\text{FP}}\approx&-\partial_N\big[1-\frac{N}{\bar{n}_\mathrm{ex}}\big]N
+\partial_\phi\big[\tilde{\Delta}-2\tilde{\eta}\sin(2\phi)\big]\\
&+\frac{3}{2}\partial_N N\partial_N+\frac{3}{8N}\partial_\phi^2.
\end{split}
\end{equation}
Furthermore, for large $\bar{n}_{\mathrm{ex}}$ and small $|\tilde{\Delta}|,\tilde{\eta}$, the intensity dynamics is much faster than the phase dynamics [see for instance Ref. \cite{BenArosh2021} for a similar discussion in a closely related system], and we make the approximation of linearizing the intensity dynamics around its stable value. Such a linearized description is obtained by defining the intensity fluctuations:
\begin{equation}\label{intensity_flucts}
\delta N=\bar{N}-\bar{n}_\mathrm{ex},
\end{equation}
and substituting this in Eq. (\ref{eq_FP3}), keeping only up to linear terms. This approximation allows us to completely decouple intensity and phase fluctuations and thus to obtain an effective stochastic model for the phase alone, i.e. Eq. (\ref{Phase_Langevin}). As we will see, this effective description captures the slowest fluctuation modes of the system, i.e. those associated with the leading eigenvalues of the Liouvillian. In terms of the Fokker-Planck equation, we obtain:
\begin{equation}\label{eq_FP4}
\begin{split}
\partial_\tau W(\delta N,\phi,\tau)=(\hat{L}_{\delta N}+\hat{L}_\phi) W(\delta N,\phi,\tau),
\end{split}
\end{equation}
with
\begin{equation}\label{FP_intenisty}
 \hat{L}_{\delta N}=\partial_{\delta N}{\delta N}+\frac{3\bar{n}_\mathrm{ex}}{2}\partial_{\delta N}^2,
\end{equation}
and
\begin{equation}\label{FP_phase}
\hat{L}_\phi=\partial_\phi\big(\tilde{\Delta}-2\tilde{\eta}\sin(2\phi)\big)+\frac{3}{8\bar{n}_\mathrm{ex}}\partial_\phi^2, 
\end{equation}
This equation displays the same intensity fluctuations of the Gaussian process in the absence of squeezing while the phase is now governed by a nonlinear drift with additive Gaussian noise. Notice that this nonlinearity makes the phase dynamics non-Gaussian. Since intensity fluctuations and phase dynamics are completely decoupled, the phase dynamics can be written in terms of the  one-dimensional Langevin equation written in the main text, i.e. Eq. (\ref{Phase_Langevin}), which provides an intuitive picture of the stochastic dynamics as resulting from the interplay of a tilted washboard potential, $V(\phi)$ [Eq. (\ref{Potential_Phase})], and a Gaussian white noise term, $\xi_\phi(t)$, whose intensity decreases with the mean-field excitation number $\bar{n}_\mathrm{ex}$. The approximations performed in this section are accurate for small detuning and squeezing strengths, and they will enable us to obtain accurate results for the leading eigenvalues of the full model. Finally, we recall that  while here we have heuristically discussed these approximations, a more rigorous treatment for this and other related systems can be found in Ref. \cite{Kato2019}. 

\subsection{Eigenspectrum of the Fokker-Planck operator.}\label{sec_A_3}

In this section, we  briefly review how a Fokker-Planck equation can be treated by eigenvalue methods (for a thorough introduction see Ref. \cite{RiskenBook}) and we write down some general properties of the eigenvalue problem associated with the Fokker-Planck operator of Eq. (\ref{eq_FP4}).

We consider the following Fokker-Planck equation
\begin{equation}
\partial_t W(\{x\},t)=\hat{L}_\mathrm{FP}W(\{x\},t),
\end{equation}
for the probability distribution $W(\{x\},t)$ and where $\{x\}$ is the set of variables of the problem. This operator is defined by both the Fokker-Planck equation and the boundary conditions. The boundary conditions for Eq. (\ref{eq_FP1}) are the so-called {\it natural boundary conditions}, in which $W(\alpha,\alpha^*,t)$ is a normalizable function that vanishes at infinity, together with the associated probability current \cite{GardinerBook,RiskenBook}. The operator $\hat{L}_\mathrm{FP}$ is generally non-Hermitian. Thus, similar to the Liouvillian, it is not necessarily diagonalizable, e.g., there may be exceptional points at which this operator can only be reduced to a Jordan normal form \cite{RiskenBook}. Away from these points, we can decompose the dynamics of the probability distribution in terms of the eigenspectrum of $\hat{L}_\mathrm{FP}$. In particular, we define its right and left eigenfunctions as:
\begin{equation}
\begin{split}
\hat{L}_\mathrm{FP}\Psi_n(\{x\})=\tilde{\lambda}_n \Psi_n(\{x\}),\\ \hat{L}^\dagger_\mathrm{FP}\bar{\Psi}^\dagger_n(\{x\})=\tilde{\lambda}_n \bar{\Psi}^\dagger_n(\{x\}),
\end{split}
\end{equation}
from which we can form a biorthonormal basis:
\begin{equation}
%\langle \bar{\Psi}^\dagger_n(\{x\}), \Psi_m(\{x\})\rangle=
\int d\{x\} \bar{\Psi}^\dagger_n(\{x\}) \Psi_m(\{x\})=\delta_{n,m}.
\end{equation}
Then, we obtain:
\begin{equation}
\begin{split}
W(\{x\},t)=\sum_n A_n \Psi_n(\{x\}) e^{\tilde{\lambda}_n t}, \\
A_n=\int d\{x'\} \bar{\Psi}^\dagger_n(\{x'\})W(\{x'\},t=0).
\end{split}
\end{equation}
This formula is analogous to the one used in the Liouvillian formalism for the decomposition of the dynamics in terms of the Liouvillian eigenmodes. Since the probability distribution $W(\{x\},t)$ is real-valued, the eigenvalues must be real or appear in complex conjugate pairs. Moreover, if the drift and diffusion terms of $\hat{L}_\mathrm{FP}$ have no singularities, the diffusion is always positive, and $\hat{L}_\mathrm{FP}$ is time-independent, there is a stationary state corresponding to:
\begin{equation}
\Psi_0(\{x\})=W_\mathrm{ss}(\{x\}),\quad \bar{\Psi}^\dagger_0(\{x\})=\mathbb{I},\quad \lambda_0=0, 
\end{equation}
while the real part of the rest of the eigenvalues is non-positive \cite{RiskenBook}.

We now consider the particular case of Eq. (\ref{eq_FP4}) in which we work with an effective description where intensity and phase dynamics are decoupled. Because of this decoupling, i.e. because of the separability of Eq. (\ref{eq_FP4}),  we can rewrite the Fokker-Plank equation as:
\begin{equation}
W(\delta N,\phi,\tau)=P(\delta N,\tau)Q(\phi,\tau), 
\end{equation}
where
\begin{equation}
\begin{split}
\partial_\tau P(\delta N,\tau)&=\hat{L}_{\delta N} P(\delta N,\tau),\\
\partial_\tau  Q(\phi,\tau)&=\hat{L}_\phi Q(\phi,\tau).
\end{split}
\end{equation}
Therefore, the complete eigenvalue problem can be decomposed into two independent one-dimensional eigenvalue problems. Defining:
\begin{equation}\label{eig_problem_specific}
\begin{split}
\hat{L}_{\delta N} \psi_n(\delta N)=\mu_n\psi_n(\delta N), \\
\hat{L}_\phi \chi_m(\phi)=\nu_m\chi_m(\phi),
\end{split}
\end{equation}
then:
\begin{equation}
\begin{split}
\hat{L}_{\text{FP}}\Psi_{n,m}(\delta N,\phi)&=\tilde{\lambda}_{n,m}\Psi_{n,m}(\delta N,\phi), \\  \Psi_{n,m}(\delta N,\phi)&=\psi_n(\delta N)\chi_m(\phi),\\
\tilde{\lambda}_{n,m}&=\mu_n+\nu_m.
\end{split}
\end{equation}
We analyze the intensity and phase eigenvalue problems separately in Secs.  \ref{App_inetnsity_eigs} and \ref{App_phase_eigs}, respectively. 

\section{Analysis of the intensity eigenvalues}\label{App_inetnsity_eigs}

In this section, we analyze the eigenspectrum of the intensity fluctuations as described by $\hat{L}_{\delta N}$ [Eq. (\ref{FP_intenisty})]. This corresponds to an Ornstein-Uhlenbeck process in the domain $\delta N\in(-\bar{n}_\mathrm{ex},\infty)$, since the total intensity $N=\bar{n}_\mathrm{ex}+{\delta N}$ is positive definite. In order to obtain analytical results we make the simplification of relaxing the finite range of ${\delta N}$  and consider  $\delta N\in(-\infty,\infty)$. This is a reasonable approximation as this variable models the fluctuations around a large mean value and thus $|\delta N|\ll\bar{n}_\mathrm{ex}$ should hold. If the solution obtained had a significant probability for values of $\delta N$ that are not small compared to $\bar{n}_\mathrm{ex}$, then the linearization procedure would not be correct. Considering this expanded range and natural boundary conditions the problem is analytically solvable (see e.g. \cite{GardinerBook,RiskenBook}). The eigenvalues are:
\begin{equation}\label{eigs_intensity}
\mu_n=-n \quad n=0,1,2,3,\dots 
\end{equation}
and thus proportional to unity (to $\gamma_1$ when considering the bare time $t$). This means that they converge to a finite value with increasing $\bar{n}_\mathrm{ex}$, and hence, the fluctuation spectrum of the intensity is gaped. Moreover, the right and left eigenfunctions read:
\begin{equation}
\begin{split}
\psi_n(\delta N)&=\sqrt{\frac{w}{ 2^n n! \pi  }}e^{-w \delta N^2}H_n(\delta N\sqrt{w}),\\ \bar{\psi}^\dagger_n(\delta N)&=\frac{1}{\sqrt{ 2^n n!}}H_n(\delta N\sqrt{w}), \quad w= \frac{1}{3\bar{n}_\mathrm{ex}},
\end{split}
\end{equation}
where $H_n(x)$ are the Hermite polynomials, e.g. $H_0(x)=1$, $H_1(x)=2x$, $H_2(x)=4x^2-2$ \footnote{Notice that these polynomials satisfy the (Gaussian) weighted orthogonality condition $\int_{-\infty}^\infty dx H_n(x)H_m(x)e^{-x^2}=\sqrt{\pi}2^n n! \delta_{n,m}.$}. Interestingly, the stationary state can be written as:
\begin{equation}
P_{\mathrm{ss}}(\delta N)=\psi_0(\delta N)=\frac{1}{\sqrt{3\pi \bar{n}_\mathrm{ex}}}e^{-\frac{1}{3\bar{n}_\mathrm{ex}}\delta N^2}. 
\end{equation}
This defines a Gaussian distribution with standard deviation $\sqrt{3\bar{n}_\mathrm{ex}/2}$. Thus, the ratio of the standard deviation of the intensity fluctuations over the mean-field intensity vanishes in the classical limit.

\section{Analysis of the phase eigenvalues}\label{App_phase_eigs}

In this appendix, we analyze the eigenspectrum of the phase dynamics as described by the phase Fokker-Planck operator $\hat{L}_\phi$  [Eq. (\ref{FP_phase})] with periodic boundary conditions. As $\phi$ is an angular variable and we do not distinguish between full rotations, the physical space is restricted to $\phi \in [0,2\pi)$. 
A practical way to automatically incorporate the boundary conditions is to write the eigenfunctions in terms of Fourier modes \cite{RiskenBook}:
\small
\begin{equation}
\begin{split}
\chi_n(\phi)=\frac{1}{2\pi}\sum_{q=-\infty}^\infty c_{q,n}e^{iq\phi},
\end{split}
\end{equation}
\normalsize
where $q$ takes integer values. Introducing this expansion in the phase eigenvalue problem [Eq. (\ref{eig_problem_specific})] leads to the following infinite recurrence relation from which all eigenvalues and eigenfunctions can be determined:
\begin{equation}\label{recc0}
\big[i\tilde{\Delta} q-\frac{3q^2}{8\bar{n}_\mathrm{ex}}-\nu_n\big]c_{q,n}-\tilde{\eta} \,q\,c_{q-2,n}+\tilde{\eta}\, q\,c_{q+2,n}=0.
\end{equation}
Importantly, the $c_q$'s with odd $q$ are decoupled from those with even $q$. This  means that the operator $\hat{L}_\phi$ can be separated into two independent 'blocks' depending on the parity of the eigenfunctions, as functions that are only made of even (odd) Fourier modes are symmetric (antisymmetric) under a $\phi\to\phi+\pi$ translation. Therefore, defining:
\begin{equation}
a_{q,n}=c_{2q,n},\quad b_{q,n}=c_{2q+1,n},\quad q=0,\pm1,\pm2,\dots . 
\end{equation}
we obtain two independent recurrence relations from the original one (\ref{recc0}):
\begin{equation}\label{recc1}
\big[i2\tilde{\Delta} q-\frac{3 q^2}{2\bar{n}_\mathrm{ex}} -\nu_{a,n}\big]a_{q,n}-2\tilde{\eta}\, q \,a_{q-1,n}+2\tilde{\eta}\, q\, a_{q+1,n}=0,
\end{equation}
\begin{equation}\label{recc2}
\begin{split}
\big[i\tilde{\Delta}(2q+1)-&\frac{3(2q+1)^2}{8\bar{n}_\mathrm{ex}} -\nu_{b,n}\big]b_{q,n}\\
&-\tilde{\eta} (2q+1) b_{q-1,n}+\tilde{\eta} (2q+1) b_{q+1,n}=0.
\end{split}
\end{equation}
Notice that we have introduced the new subindex $a,b$ to distinguish the eigenvalues of the two symmetry sectors $\nu_{a(b)}$ and the corresponding eigenfunctions:
\begin{equation}
\begin{split}
\chi_{a,n}(\phi)&=\frac{1}{2\pi}\sum_{q=-\infty}^\infty a_{q,n}e^{i 2q\phi}, \\ \chi_{b,n}(\phi)&=\frac{1}{2\pi}\sum_{q=-\infty}^\infty b_{q,n}e^{i (2q+1)\phi},
\end{split}
\end{equation}
from which their symmetry/antisymmetry is evident. This can be understood as the manifestation of the parity symmetry of the model, i.e. the invariance of Eq. (\ref{Wigner_ME}) under the transformation $\alpha\to-\alpha$ and $\alpha^*\to-\alpha^*$ which has not been lost in the subsequent approximations. Moreover, the stationary state belongs to the symmetric sector, i.e. 'a'.

Finally, in order to obtain a solution for these infinite recurrence relations one must generally resort to truncation \cite{RiskenBook}. Then we obtain a finite recurrence relation that can be mapped  to two independent matrices, recovering the standard matrix eigenvalue problem. Thus, considering a truncation at $q=M$ (for $q=-M$ is analogous) the recurrence relations are closed as:
\begin{equation}
\big[i2\tilde{\Delta} M-\frac{3 M^2}{2\bar{n}_\mathrm{ex}}-\nu_{a,n}\big]a_{M,n}-2\tilde{\eta}\, M \,a_{M-1,n}=0,
\end{equation}
and
\small
\begin{equation}
\big[i\tilde{\Delta}(2M+1)-\frac{3(2M+1)^2}{8\bar{n}_\mathrm{ex}} -\nu_{b,n}\big]b_{M,n}-\tilde{\eta} (2M+1) b_{M-1,n}=0.
\end{equation}
\normalsize
The results are meaningful as long as they do not vary significantly when considering the next truncation size: $M+1$. In this sense, this method parallels the usual truncation schemes used to numerically diagonalize (e.g. bosonic) Liouvillians. 

\subsection{Limiting case of $\eta=0$.} 

This is an illustrative case in which the phase dynamics can  be analytically solved. Moreover, this will allow us to show explicitly how the decay rate of the phase eigenvalues are exactly proportional to $1/\bar{n}_\mathrm{ex}$, and thus they become purely imaginary in the infinite excitation limit. 

Fixing $\eta=0$ in the first recurrence relation (\ref{recc0}) yields:
\begin{equation}%\label{recc0}
\big[i\tilde{\Delta} q-\frac{3q^2}{8\bar{n}_\mathrm{ex}}-\nu_n\big]c_{q,n}=0.
\end{equation}
Hence, the different Fourier modes are not coupled by the phase dynamics, and they are actually the eigenmodes of the problem for $\eta=0$. The eigenspectrum in this case is:
\begin{equation}\label{eigs_phase_free}
\nu_n=i\tilde{\Delta} n-\frac{3n^2}{8\bar{n}_\mathrm{ex}}, \quad n=0,\pm1,\pm2,\dots 
\end{equation}
\begin{equation}
 \chi_n(\phi)=\frac{1}{2\pi} e^{in\phi},\quad  \bar{\chi}^\dagger_n(\phi)=e^{-in\phi},
\end{equation}
from which we can easily check that $\int_0^{2\pi}d\phi \chi_n(\phi)\bar{\chi}^\dagger_m(\phi)=\delta_{n,m}$. In this case the stationary state corresponds to  the uniform phase distribution $Q_\mathrm{ss}=1/(2\pi)$. 

\subsection{Phase-difusion regime}\label{App_phase_difusion}

For non-zero squeezing strength, the recurrence relations given in Eqs. (\ref{recc1}) and (\ref{recc2}) cannot be analytically solved and one must resort to numerical diagonalization of the matrix representation of their truncated form. Nevertheless, in the limit-cycle regime $\eta<\eta_\mathrm{c}$, we can use perturbation theory to show that the decay rate of the leading phase eigenmodes scales at most as $1/\bar{n}_\mathrm{ex}$, such that in the infinite-excitation limit they become purely imaginary. We proceed by focusing on the eigenvalues of the odd-parity sector, to which the two leading excitation eigenmodes belong.  Afterwards, we comment on the extension to the even-parity sector which follows similar lines.

The starting point is to rewrite the (truncated) recurrence series  Eq. (\ref{recc2}) in matrix form, which we denote as $\Phi^{\mathrm{(b)}}$. The  matrix elements read:
\small
\begin{equation}
\begin{split}
\Phi^{\mathrm{(b)}}_{j,k}=&i\tilde{\Delta}\big[2\big(j-(M+1)\big)+1\big]\delta_{j,k}\\
&-\frac{3}{8\bar{n}_\mathrm{ex}} \big[2\big(j-(M+1)\big)+1\big]^2\delta_{j,k}\\
&-\tilde{\eta} [2\big(j-(M+1)\big)+1]\delta_{j,k-1}(1-\delta_{j,0})\\
&+\tilde{\eta} [2\big(j-(M+1)\big)+1]\delta_{j,k+1}(1-\delta_{j,2M+1}),
\end{split}
\end{equation}
\normalsize
where $j,k\in[0,2M+1]$, $\delta_{j,k}$ is the Kronecker delta and $M$ is the truncation dimension of the originally infinite recurrence relation given in Eq. (\ref{recc2}). Notice that from  Eq. (\ref{recc1}), we can define the corresponding matrix for the even-parity sector, which in this case splits into two independent matrices for positive and negative $q's$. These are complex conjugates of each other and we denote them by $\Phi^{\mathrm{(a)}}$. These matrices contain the stationary state and the even-parity excitation spectrum. The matrices $\Phi^{\mathrm{(a,b)}}$ are non-Hermitian and tridiagonal. We now split $\Phi^{\mathrm{(b)}}$ in two matrices:
\begin{equation}\label{structure_phase_eigs}
\Phi^{\mathrm{(b)}}=\mathcal{H}+\frac{1}{\bar{n}_\mathrm{ex}} \mathcal{V}
\end{equation}
with
\begin{equation}
\begin{split}
\mathcal{H}_{j,k}=&i\tilde{\Delta}\big[2(j-(M+1))+1\big]\delta_{j,k}\\
&-\tilde{\eta} (2(j-(M+1))+1)\delta_{j,k-1}(1-\delta_{j,0})\\
&+\tilde{\eta} (2(j-(M+1))+1)\delta_{j,k+1}(1-\delta_{j,2M+1}),
\end{split}
\end{equation}
and 
\begin{equation}
\begin{split}
\mathcal{V}_{j,k}=-\frac{3}{8}\big[2(j-(M+1))+1\big]^2\delta_{j,k}
\end{split}
\end{equation}
Notice that $\mathcal{H}$ contains the information about the mean-field dynamics while $\mathcal{V}$ introduces the effects of diffusion. In these terms, $\mathcal{H}$ only depends on $\tilde{\Delta}$ and $\tilde{\eta}$, while $\mathcal{V}$ is {\it independent} of the parameters of the system, and $\bar{n}_\mathrm{ex}$ enters as a constant dividing $\mathcal{V}$. This form of rewriting the problem suggests a perturbative approach for the classical limit (i.e. for $\bar{n}_\mathrm{ex}\to \infty$). One can numerically check that the eigenvalues of $\mathcal{H}$ are given by $i(2n+1)\Omega$ with $n=0,\pm1,\pm2,\dots$ in a very good approximation (related to the finite truncation of the recurrence series). Therefore, the separation between adjacent eigenvalues is $2\Omega$, and for $(1/\bar{n}_\mathrm{ex})\ll\Omega$, we should be able to account for the effects of $\mathcal{V}$ perturbatively. In particular, we will restrict ourselves to the first-order corrections to the eigenvalues, as we will show they provide us the scaling of the lifetime with the mean-field excitation number $\bar{n}_\mathrm{ex}$. 

The starting point of such perturbative treatment is the diagonalization of $\mathcal{H}$:
\begin{equation}
\mathcal{H}|R^{(0)}_j\rangle=\nu^{(0)}_j  |R^{(0)}_j\rangle, \quad  \langle L^{(0)}_j|\mathcal{H}=\nu^{(0)}_j  \langle L^{(0)}_j|
\end{equation}
where the right and left eigenvectors form a biorthogonal system that can be normalized in such a way that $\langle L^{(0)}_j|R^{(0)}_k\rangle=\delta_{j,k}$. Next, we consider the following perturbation series:
\small
\begin{equation}
\begin{split}
&(\mathcal{H}+\frac{1}{\bar{n}_\mathrm{ex}} \mathcal{V})\big(|R^{(0)}_j\rangle+\frac{1}{\bar{n}_\mathrm{ex}}|R^{(1)}_j\rangle +\frac{1}{\bar{n}^2_\mathrm{ex}}|R^{(2)}_j\rangle+\dots \big)\\
&=(\nu^{(0)}_j+\frac{1}{\bar{n}_\mathrm{ex}}\nu_j^{(1)}+\frac{1}{\bar{n}^2_\mathrm{ex}}\nu_j^{(2)}\dots)\\
&\cdot\big(|R^{(0)}_j\rangle+\frac{1}{\bar{n}_\mathrm{ex}}|R^{(1)}_j\rangle +\frac{1}{\bar{n}^2_\mathrm{ex}}|R^{(2)}_j\rangle+\dots \big),
\end{split}
\end{equation}
\normalsize
where the superscript denotes the perturbation order of the expansion in terms of $1/\bar{n}_\mathrm{ex}$. Separating the zero and first orders in $1/\bar{n}_\mathrm{ex}$ we obtain:
\begin{equation}
\mathcal{H}|R^{(0)}_j\rangle=\nu^{(0)}_j|R^{(0)}_j\rangle,
\end{equation}
and
\begin{equation}
\mathcal{H}|R_j^{(1)}\rangle+\mathcal{V}|R^{(0)}_j\rangle=\nu^{(0)}_j |R_j^{(1)}\rangle+ \nu^{(1)}_j |R^{(0)}_j\rangle.
\end{equation}
We multiply on the left by $\langle L^{(0)}_j|$, we use the biorthogonality property and that $\langle L^{(0)}_j|$ is an eigenvector of $\mathcal{H}$,  obtaining:
\begin{equation}\label{first_order}
\nu_j^{(1)}=\langle L^{(0)}_j|\mathcal{V}| R^{(0)}_j\rangle.     
\end{equation}
Since $|R^{(0)}_j\rangle$, $\langle L^{(0)}_j|$ are not Hermitian conjugates, we can not tell in advance whether $\nu_j^{(1)}$ is real or complex. Thus, in principle we have two possibilities: (i) the first-order correction is real-valued, leading to a finite decay rate with a scaling $1/\bar{n}_\mathrm{ex}$; (ii) the first-order correction is complex-valued, leading additionally to a frequency correction of order $1/\bar{n}_\mathrm{ex}$. Notice that by the properties of the Fokker-Planck equation, the possible real part of the eigenvalues must be negative \cite{RiskenBook}. In both cases, the conclusion is that in the perturbative regime, the eigenvalues can acquire at most a decay rate that scales inversely proportional to the mean-field excitation number. Indeed, we have taken a step forward and we have obtained numerically Eq. (\ref{first_order}) for parameter values and different eigenmodes, always finding that it is real and negative, which suggests that this might be the case for all eigenmodes and in all the limit-cycle regime.  Therefore, to first order in $1/\bar{n}_\mathrm{ex}$, the eigenvalues of $\Phi^{\mathrm{(b)}}$ are given by:
\begin{equation}\label{odd_eigvals}
\nu_{b,n}=i(2n+1)\tilde{\Omega} -\frac{1}{\bar{n}_\mathrm{ex}} c_{2n+1}, \quad n=0,\pm1,\pm2,\dots   
\end{equation}
where $c_{2n+1}$ is a positive real constant that depends on the eigenmode and on $\Omega$, according to Eq. (\ref{first_order}). Numerically we observe that $c_n=c_0 n^2$, similarly to the exact analytical results for $\eta=0$. This provides an argument for the  scaling of the decay rates with  $1/\bar{n}_\mathrm{ex}$ for large enough $\bar{n}_\mathrm{ex}$. Interestingly, as $\Omega$ vanishes at the bifurcation point, this means that the perturbative regime in which the first order results are accurate occurs for ever increasing $\bar{n}_\mathrm{ex}$ as the bifurcation is approached. This also provides an argument to explain why the linear regime needs larger $\bar{n}_\mathrm{ex}$ to be observed as $\tilde{\eta}$ is increased towards the critical value (see Fig. \ref{fig_sup_eigs}).

  \begin{figure}[t!]
 \centering
 \includegraphics[width=1\columnwidth]{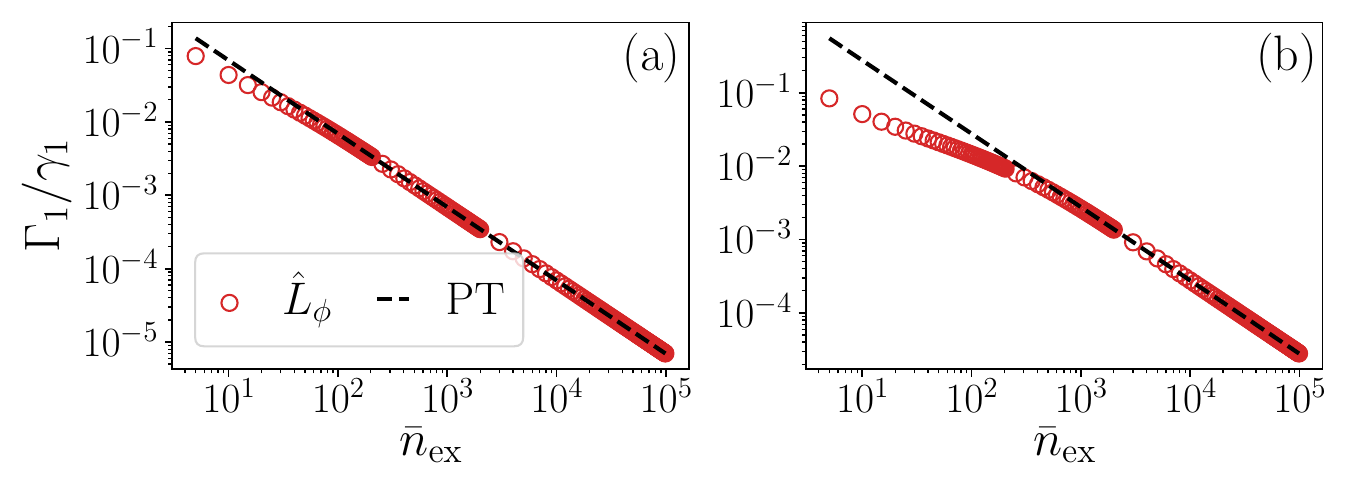}
 \caption{{\bf Emergence of the perturbative regime in phase diffusion timescales.} Red empty dots: leading decay rate of the effective phase model. Black dashed line: decay rate according to the first order perturbation theory as given in Eq. (\ref{odd_eigvals}). (a) $\eta/\eta_\mathrm{c}=0.6$, with the perturbation theory results $c_1=0.69$. (b) $\eta/\eta_\mathrm{c}=0.9$ with the perturbation theory results $c_1=2.77$. In both cases $\Delta/\gamma_1=0.1$. The comparison between (a) and (b) shows that the closer we are to the bifurcation, the larger needs to be $\bar{n}_\mathrm{ex}$ in order for the perturbative results to be accurate.}
 \label{fig_sup_eigs}
\end{figure}
 
Finally we notice that the same perturbative arguments apply to the even-parity {\it excitation} modes. This is because  $\Phi_q^{\mathrm{(a)}}$ displays essentially the same structure as in Eq. (\ref{structure_phase_eigs}), which is the basis of the perturbative treatment. The only important difference to take into account is that $\Phi_q^{\mathrm{(a)}}$ contains the stationary state and thus a zero eigenvalue. This eigenvalue and the corresponding eigenvector must be excluded from the perturbation series expansion. However, such a perturbative expansion can still be pursued for the rest of the excitation eigenmodes, thus generalizing Eq. (\ref{odd_eigvals}) to Eq. (\ref{phase_eigenvalues}) of the main text. In  conclusion, in the perturbative regime, they also acquire a decay rate that scales at maximum as $1/\bar{n}_\mathrm{ex}$, as shown in Fig. \ref{fig_continuous} (d) for the first two leading eigenmodes of the even-parity sector.

\subsection{Quantum activation regime}\label{App_quantum_activation}

In the bistable regime, the longest timescale corresponds to an activation process in which noise induces the system to jump from one stable solution to the other. This timescale is given by the inverse of the rate of jumps, which we calculate in this appendix.

Due to the periodicity of the potential we can analyze this process by initially considering just one of the stable fixed points. Each fixed point has a potential barrier on its right and on its left (see Fig. \ref{fig_potential} for a graphical guide). The rate of left or right transitions corresponds to the inverse of the escape time through the left or right potential barrier \cite{GardinerBook,RiskenBook}. In the small noise approximation (small diffusion constant compared to potential barrier height), the well-known Kramer's escape rate formula can be used  \cite{GardinerBook,RiskenBook}:
\begin{equation}
\Gamma_\mathrm{esc}=\frac{1}{2\pi}\sqrt{|V''(\phi_\mathrm{M})|V''(\phi_\mathrm{m})}\exp\big[\frac{V(\phi_\mathrm{m})-V(\phi_\mathrm{M})}{D}\big], 
\end{equation}
where $\phi_\mathrm{M}$ denotes the phase at one of the local maxima of the potential defined in Eq. (\ref{Potential_Phase}), while $\phi_\mathrm{m}$ the phase at the potential minimum corresponding to one of the stable fixed points. Derivatives with respect to $\phi$ are denoted with a prime, while $D$ is the diffusion constant, in our model given by $D=3/(8\bar{n}_\mathrm{ex})$ [Eq. (\ref{FP_phase})]. Particularizing this formula for the potential barrier at the right of a stable fixed point we obtain:
\small
\begin{equation}\label{gamma_right}   
\frac{\Gamma_\rightarrow}{\gamma_1}=\frac{\tilde{C}}{2} \exp\bigg[-\frac{8\bar{n}_\mathrm{ex}}{3}\bigg(\sqrt{4\tilde{\eta}^2-\tilde{\Delta}^2}+\tilde{\Delta}\sin^{-1}\frac{|\tilde{\Delta}|}{2\tilde{\eta}}+\frac{\tilde{\Delta} \pi}{2}\bigg)\bigg]    
\end{equation}
\normalsize
with
\begin{equation}
\tilde{C}=\frac{2}{\pi}\sqrt{4\tilde{\eta}^2-\tilde{\Delta}^2}    
\end{equation}
while for the left potential barrier it leads to:
\small
\begin{equation}\label{gamma_left}   
\frac{\Gamma_\leftarrow}{\gamma_1}= \frac{\tilde{C}}{2}  \exp\bigg[-\frac{8\bar{n}_\mathrm{ex}}{3}\bigg(\sqrt{4\tilde{\eta}^2-\tilde{\Delta}^2}+\tilde{\Delta}\sin^{-1}\frac{|\tilde{\Delta}|}{2\tilde{\eta}}-\frac{\tilde{\Delta} \pi}{2}\bigg)\bigg] 
\end{equation}
\normalsize
Taking their ratio we see that jumps to the right (or left) are exponentially suppressed with respect to jumps to the left (or right) when the detuning is positive (negative). For this reason, we take into account only those occurring through the dominant direction. Then, taking into account that there are two stable fixed points and that they are connected through periodic boundary conditions, we identify the Liouvillian gap as two times the dominant direction rate \cite{RiskenBook}:
\begin{equation}
\Gamma_\mathrm{gap}=2\Gamma_{\leftarrow}, \quad \text{for} \quad \Delta > 0 
\end{equation}
and the other way around for $\Delta<0$. This result should be accurate for a large number of excitations and well into the bistable regime. Notice that we do not expect this formula to work near the bifurcation as there the noise intensity becomes relatively large compared to the potential barrier which tends to be flat. Analogously to the limit-cycle regime, a given noise intensity, or the number of excitations, determines the notion of close or far from the bifurcation. Finally,  taking the absolute value of the detuning, we arrive at the formula given in the main text:
\small
\begin{equation}
\frac{\Gamma_\mathrm{gap}}{\gamma_1}=\tilde{C} \exp\bigg[-\frac{8\bar{n}_\mathrm{ex}}{3}\bigg(\sqrt{4\tilde{\eta}^2-\tilde{\Delta}^2}+\tilde{\Delta}\sin^{-1}\frac{|\tilde{\Delta}|}{2\tilde{\eta}}-\frac{|\tilde{\Delta}| \pi}{2}\bigg)\bigg],  
\end{equation}
\normalsize
which we have shown to be in excellent agreement with the exact results in Fig. \ref{fig_parity}.

\section{Supplemental results for the continuous time-crystal regime}\label{app_continuous}

In this appendix, we include some supplemental results regarding continuous time-symmetry breaking. In the main text, we have focused on the behavior of the decay rates of the fundamental band of eigenmodes, which vanish in the infinite excitation limit, while their imaginary part or eigenfrequency tends to multiples of the mean-field one. Here, we exemplify the latter. In particular, in Fig. \ref{fig_sup_LC} (a) we plot the difference between the imaginary part of the first four modes of the fundamental band and the corresponding multiple of the mean-field frequency:
\begin{equation}\label{frequency_difference}
\Delta_n=\varepsilon_n-\Omega_n, \quad \Omega_n=n\Omega.
\end{equation}
As we can see their relative difference (weighted by $\Omega_n$) diminishes with $\bar{n}_\mathrm{ex}$ and indeed seems to tend towards zero as the mean-field excitation number increases. Notice how higher harmonics seem to be more affected by finite size effects, as their associated $\Delta_n$ needs of larger $\bar{n}_\mathrm{ex}$ to go below a certain value.

\begin{figure}[t!]
 \centering
 \includegraphics[width=1\columnwidth]{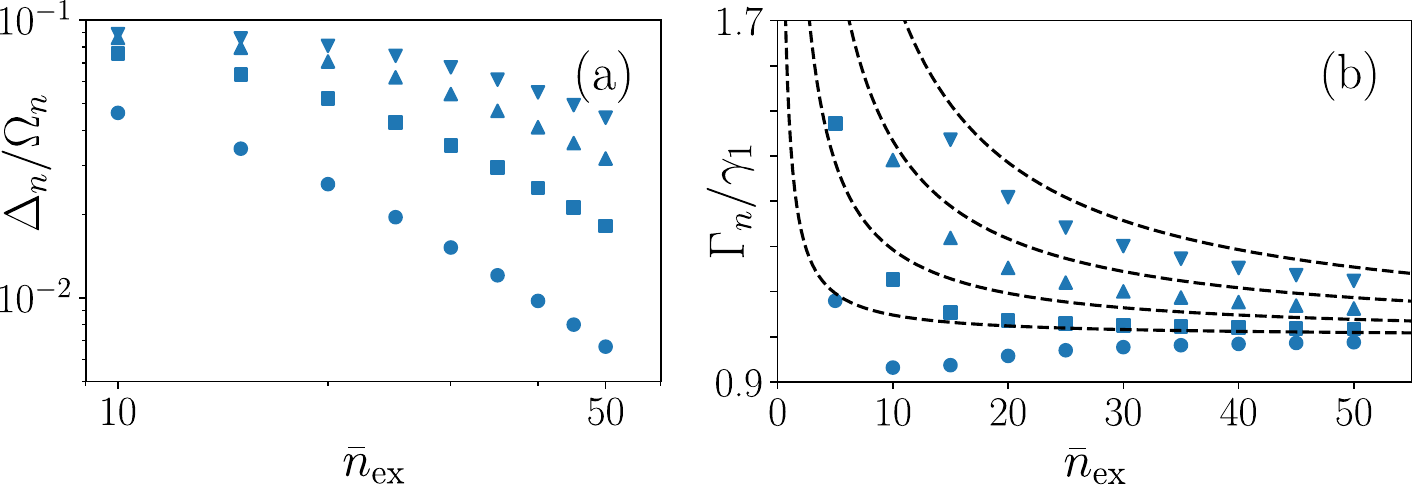}
 \caption{{\bf Supplemental results for the eigenvalues in the limit-cycle regime.} (a) Difference between the eigenfrequency of the first four modes of the fundamental band and the corresponding multiple of the mean-field frequency [Eq. (\ref{frequency_difference})], varying $\bar{n}_\mathrm{ex}$. Circles correspond to the first eigenfrequency ($n=1$), squares to the second ($n=2$), triangles to the third  ($n=3$) and inverted triangles to the fourth ($n=4$). (b) Blue markers: decay rate of the first four Liouvillian eigenvalues of the second band. Black-dashed lines correspond to the approximate expression given in Eq. (\ref{second_band})  with the same values of $c_n$ as given in Fig. \ref{fig_continuous}. In both cases $\eta/\eta_\mathrm{c}=0.4$ and $\Delta/\gamma_1=0.1$.}
 \label{fig_sup_LC}
\end{figure}

We now analyze the behavior of the decay rates of the second band of eigenmodes and we exemplify how they saturate to $\gamma_1$ with increasing $\bar{n}_\mathrm{ex}$. In Fig. \ref{fig_sup_LC} (b) we compare the exact results for the first four eigenvalues of the second band (blue markers)  with the  approximate ones (black dashed lines). The latter are obtained by combining the intensity fluctuation eigenvalue for $m=1$ [Eq. (\ref{intensity_eigenvalues})]  with the first four phase eigenvalues obtained with perturbation theory [Eq. (\ref{phase_eigenvalues})]:
\begin{equation}\label{second_band}
\text{Re}[\tilde{\lambda}_{1,n}]\approx-\gamma_1-\frac{\gamma_1 c_n}{\bar{n}_\mathrm{ex}}, \quad n=1,2,3,4.
\end{equation}
From this figure, we observe that the approximate results capture reasonably well the behavior of the exact Liouvillian eigenvalues, despite the considered $\bar{n}_\mathrm{ex}$ not being very large. Importantly, we observe that the decay rate of the second band of eigenmodes tends to saturate to $\gamma_1$, as we have commented in the main text.

\section{Supplemental results for the parity broken phase}\label{app_parity}

\begin{figure}[t!]
\includegraphics[width=1\linewidth]{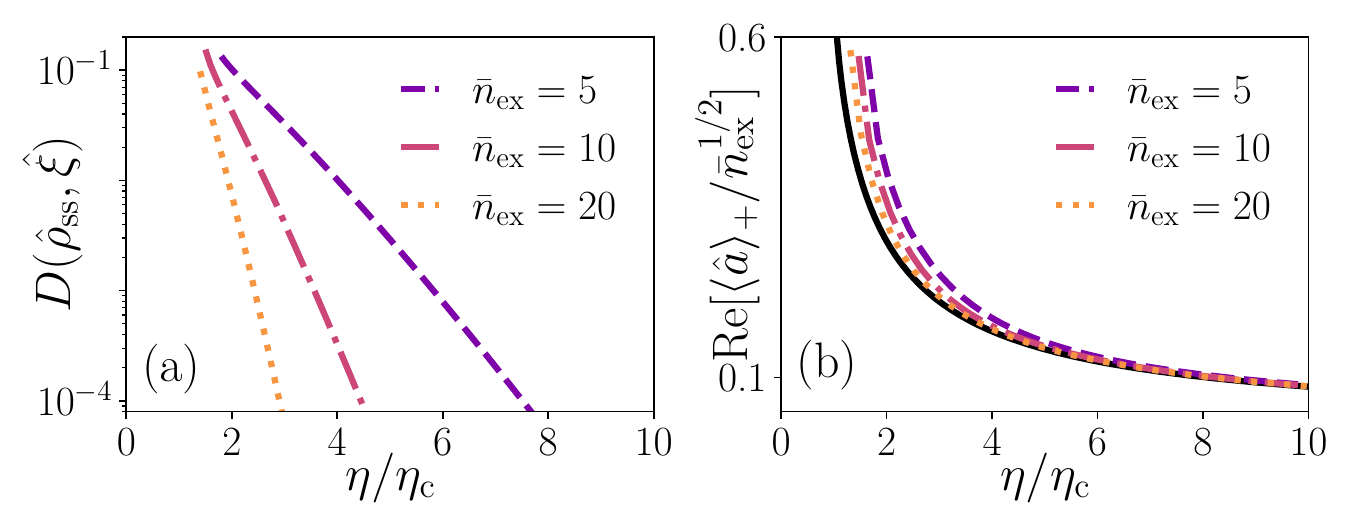}\\
\caption{{\bf Emergent symmetry broken states.} (a) Trace distance between the exact stationary state and $\hat{\xi}$. This figure of merit is plotted only for $\eta\geq\eta_\mathrm{EP}$ as it is where it is valid to decompose $\hat{r}_1$ in terms of the density matrices $\hat{\rho}_\pm$. Both axes are in log scale. (b) Comparison of the real part of $\langle \hat{a}\rangle_+$ with the (rescaled) mean-field solution in the bistable regime (black solid line).  In both panels $\Delta/\gamma_1=0.1$}\label{fig_tracedist}
\end{figure}

In this appendix, we analyze in more detail the emergence of the symmetry-broken states in the infinite excitation limit and for $\eta>\eta_\mathrm{c}$. Numerically, these states can be most easily computed from the spectral decomposition of $\hat{r}_1$. Recall that as $\lambda_1$ is real, $\hat{r}_1$ is Hermitian, while it is also traceless. This implies that it can be written as the subtraction of two density matrices, which as we shall see, do correspond to the symmetry-broken states \cite{Minganti2018}:
\begin{equation}\label{timecrystals_stateDEC}
\hat{r}_1=\frac{1}{2}(\hat{\rho}_+-\hat{\rho}_-).    
\end{equation}
Following Ref. \cite{Minganti2018}, in the limit in which $\lambda_1\to0$, $\hat{r}_1$ and $\hat{r}_0$ become degenerate, and we make the ansatz of  writing down $\hat{\rho}_\mathrm{ss}$ also in terms of $\hat{\rho}_\pm$:
\begin{equation}\label{timecrystals_Xi}
\hat{\xi}=\frac{1}{2}(\hat{\rho}_++\hat{\rho}_-), \quad \lim_{\bar{n}_\mathrm{ex}\to\infty}\hat{\rho}_\mathrm{ss}\to\hat{\xi}.
\end{equation}
This is a hypothesis to be checked numerically. Notice that by construction, it is consistent with the different symmetry of the stationary state and the leading eigenmode. In Fig. \ref{fig_tracedist} (a) we numerically check the hypothesis of Eq. (\ref{timecrystals_Xi}), by plotting the trace distance between $\hat{\rho}_\mathrm{ss}$ and $\hat{\xi}$, i.e.:
\begin{equation}
D(\hat{\rho}_\mathrm{ss},\hat{\xi})=\frac{1}{2}\text{Tr}\big[\sqrt{(\hat{\rho}_\mathrm{ss}-\hat{\xi})^\dagger(\hat{\rho}_\mathrm{ss}-\hat{\xi})}\big],
\end{equation}
for $\eta\geq\eta_\mathrm{EP}>\eta_\mathrm{c}$ and various values of $\bar{n}_\mathrm{ex}$. We can clearly appreciate how this distance vanishes as the Liouvillian gap closes, numerically confirming  Eq. (\ref{timecrystals_Xi}). Therefore, the fact that an eigenmode of a different symmetry sector (than the stationary state) displays a vanishing eigenvalue for $\eta>\eta_\mathrm{c}$ and $\bar{n}_\mathrm{ex}\to\infty$ results in the emergence of two stationary states, i.e. $\hat{\rho}_\pm$, that break parity symmetry in this regime, as explained in the main text.

As commented, this is intimately related to the presence of bistability at the mean-field level. In fact, we can check that observables calculated over these states tend to the mean-field results for each fixed point. This is illustrated in Fig. \ref{fig_tracedist} (b), in which we numerically show that $\langle \hat{a}\rangle_+=\text{Tr}[\hat{a}\hat{\rho}_+]$ tends to the mean-field solution $\alpha_+$ in the infinite-excitation limit and for $\eta\geq\eta_\mathrm{EP}$ \footnote{Notice that we plot these figures of merit  for $\eta\geq\eta_\mathrm{EP}$, and not $\eta\geq\eta_\mathrm{c}$, as for finite $N$ it is only in the former case where decomposition (\ref{timecrystals_stateDEC}) is possible. Of course, this range tends to $\eta\geq\eta_\mathrm{c}$ when $\bar{n}_\mathrm{ex}\to \infty$.}, and where ${\alpha}_+$ is one of the mean-field fixed points [see Eq. (\ref{MF_solution})]. As the lifetime of these states diverges in this limit, $\langle \hat{a}\rangle_\pm\neq0$ can be regarded as an order parameter for parity symmetry breaking \cite{Minganti2018}.  Finally, it is interesting to notice that as $\bar{n}_\mathrm{ex}$  is increased, the symmetry-broken states $\hat{\rho}_\pm$ become equivalent to the extreme metastable states that we disclosed in Ref. \cite{Cabot2021}, both descriptions being equivalent in this limit. 

\bibliography{references}

\end{document}